\documentclass[12pt,a4paper]{article}
\usepackage{graphicx}
\usepackage{dcolumn}

\begin{document}
\title{Light-ion production in the interaction\\
of 96 MeV neutrons with oxygen}
\author{ U. Tippawan$^{1,2}$, 
S. Pomp$^{1}$\footnote{Corresponding author, Tel. +46 18 471 6850, 
Fax. +46 18 471 3853, E-mail: stephan.pomp@tsl.uu.se}, 
A. Ata\c{c}$^{1}$, B. Bergenwall$^{1,4}$, J. Blomgren$^1$, \\
S. Dangtip$^{1}$, A. Hildebrand$^1$, C. Johansson$^1$, J. Klug$^{1}$, P. 
Mermod$^1$, \\
L. Nilsson$^{1,3}$, M. \"Osterlund$^1$, N. Olsson$^{1,5}$, A.V. Prokofiev$^3$,  \\
P. Nadel-Turonski$^{4}$, V. Corcalciuc$^{6}$, and A. J. Koning$^{7}$ \\
\\
\fontsize{9}{12}
  $^{1}$\textit{Department of Neutron Research, Uppsala University, Sweden} \\
\fontsize{9}{12}
  $^{2}$\textit{Fast Neutron Research Facility, Chiang Mai University, Thailand} \\
\fontsize{9}{12}
  $^{3}$\textit{The Svedberg Laboratory, Uppsala University, Sweden} \\
\fontsize{9}{12}
  $^{4}$\textit{Department of Radiation Sciences, Uppsala University, Sweden} \\
\fontsize{9}{12}
  $^{5}$\textit{Swedish Defence Research Agency, Stockholm, Sweden} \\
\fontsize{9}{12}
  $^{6}$\textit{Institute of Atomic Physics, Heavy Ion Department, Bucharest, Romania} \\
\fontsize{9}{12}
  $^{7}$\textit{Nuclear Research and Consultancy Group, Petten, The Netherlands}} 

\date{}
\maketitle
\section*{Abstract}
Double-differential cross sections for light-ion (p, d, t, $^3$He and $\alpha$) 
production in 
oxygen induced by 96 MeV neutrons are reported. Energy spectra are measured at 
eight laboratory angles from $20^\circ$ to $160^\circ$ in steps of $20^\circ$. 
Procedures for data 
taking and data reduction are presented. Deduced energy-differential and 
production cross sections are reported. Experimental cross sections are compared 
to theoretical reaction model calculations and experimental data at lower 
neutron energies in the literature. The measured proton data agree reasonably well 
with the results of the model calculations, whereas the agreement for the other
particles is less convincing. The measured production cross sections for protons,
deuterons, tritons and alpha particles support the trends suggested by data at 
lower energies.
\\
PACS numbers: 24.10.-i, 25.40.-h, 25.40.Hs, 25.40.Kv, 28.20.-v



\section{Introduction}
\label{sec:Introduction}


Fast-nucleon induced 
reactions provide useful means to investigate nuclear structure, to characterize reaction mechanisms 
and to impose stringent constraints on nuclear model calculations. Although 
oxygen is a light nucleus with doubly closed shells, it can be expected that 
many statistical assumptions hold for nucleon-induced reactions at several tens 
of MeV, because the level density at high excitation energies is  sufficiently 
high that shell effects and other nuclear structure signatures are washed out. 
Light nuclei also have low Coulomb barrier, implying that the
suppression of charged-particle emission is weak. Therefore, nuclear reaction 
models for equilibrium and pre-equilibrium decay can be tested and 
benchmarked. Experimental data on reactions in oxygen in the literature at 
incident neutron energies of 27, 40, and 60 MeV~\cite{Nee82,Sub86} and between 25 and 
65 MeV~\cite{Ben98a,Ben98b,Ben98c} offer possibilities to test the 
predictions of reaction models.

In recent years, an increasing number of applications involving fast neutrons 
have been developed or are under consideration, e.g., radiation treatment of 
cancer~\cite{Ore98,Sch01,Lar95}, neutron dosimetry at commercial aircraft
altitudes~\cite{Bar00}, soft-error effects in computer memories~\cite{Single,Cha99}, 
accelerator-driven transmutation of nuclear waste and energy 
production~\cite{High,Hind}, and determination of the response of neutron 
detectors~\cite{Cec79}. 
Data on light-ion production in light nuclei such as carbon, nitrogen and oxygen 
are particularly important in calculations of dose distributions in human tissue 
for radiation therapy at neutron beams, and for dosimetry of
high energy neutrons produced  by high-energy cosmic radiation interacting 
with nuclei (nitrogen and oxygen) in the atmosphere~\cite{Bar00,Blo03}. When studying 
neutron dose effects in radiation therapy and at high altitude, it is 
especially important to consider oxygen, because it is the dominant element 
(65 \% by weight) in average human tissue.  

In this paper, we present experimental double-differential cross sections 
(inclusive yields) for protons,
deuterons, tritons, $^3$He and alpha particles produced by 96 MeV neutrons 
incident on oxygen. The measurements 
have been performed at the cyclotron of The Svedberg Laboratory (TSL), Uppsala, 
using the 
MEDLEY experimental setup~\cite{Dan00}. Spectra have been measured at 8 
laboratory angles, ranging from 
$20^\circ$ to $160^\circ$ in $20^\circ$ steps. Extrapolation procedures are used 
to obtain coverage of 
the full angular range, and consequently energy-differential and 
production cross sections are 
deduced, the latter by integrating over energy and angle. The experimental data 
are compared to results 
of calculations with nuclear reaction codes and to existing experimental data 
at lower incident neutron energies.

The experimental methods are briefly discussed in Sec.~\ref{sec:Experimental 
methods} and data reduction and 
correction procedures are presented in Sec.~\ref{sec:Data reduction}. 
The theoretical framework is summarized in Sec.~\ref{sec:Theoretical models}. In 
Sec.~\ref{sec:Results and discussion}, the 
experimental results are reported and compared with theoretical and previous 
experimental data. Conclusions 
and an outlook are given in Sec.~\ref{sec:Conclusions and outlook}. 


\section{Experimental methods}
\label{sec:Experimental methods}


The experimental setup, procedures for data reduction and corrections have been 
described in detail recently~\cite{Tip04,Udo04} and therefore only brief 
summaries are given here.

The neutron beam facility at The Svedberg Laboratory (TSL) uses 
the $^7$Li(p,n)$^7$Be reaction to produce a 
quasi-monoenergetic neutron beam~\cite{Klu02}. The lithium target was 8 mm thick 
in the present experiment and enriched to 99.98 \% in $^7$Li. The 98.5$\pm$0.3 
MeV protons from the 
cyclotron impinge on the lithium target, producing neutrons with a full-energy peak 
of 95.6$\pm$0.5 MeV 
with a width of 1.6 MeV (FWHM). With a beam intensity of 5 $\mu$A, the 
neutron flux in the full-energy peak is about 5$\cdot$10$^4$ neutrons/(s$\cdot$cm$^2$) 
at the target location. The collimated neutron beam has a diameter of 
80 mm at the location of the target, where it is monitored by a thin film breakdown 
counter 
(TFBC)~\cite{Smi95}. Relative monitoring was obtained by charge integration of
the proton beam in a Faraday cup located in the proton beam dump. The two beam 
monitor readings were in agreement during the measurements. 

The charged particles are detected by the MEDLEY setup~\cite{Dan00}. It consists 
of eight three-element 
telescopes mounted inside a 100 cm diameter evacuated reaction chamber. Each 
telescope consists of two 
fully depleted $\Delta E$ silicon surface barrier detectors and a CsI(Tl) 
crystal. The thickness of the 
first $\Delta E$ detector ($\Delta E_1$) is either 50 or 60 $\mu$m, while the 
second one ($\Delta E_2$) 
is either 400 or 500 $\mu$m. They are all 23.9 mm in diameter (nominal). The 
cylindrical CsI(Tl) crystal, 
50 mm long and 40 mm in diameter, serves as the $E$ detector. 

A 22 mm diameter 500 $\mu$m thick (cylindrical) disk of SiO$_2$ is used as the 
oxygen target. For the subtraction of the silicon contribution, measurements using 
a silicon wafer having a 32$\cdot$32
mm$^2$ quadratic shape and a thickness of 303 $\mu$m are performed. 

For absolute cross section normalization, a 25 mm diameter and 1.0 mm thick 
polyethylene (CH$_2$)$_n$ 
target is used. The $np$ cross section at $20^\circ$ laboratory angle provides 
the reference cross 
section~\cite{Rah01}. Instrumental background is measured by removing the target from the 
neutron beam. It is dominated by protons 
produced by neutron beam interactions with the beam tube and reaction chamber 
material, especially at 
the entrance and exit of the reaction chamber and in the telescope housings. 
Therefore, the telescopes 
at $20^\circ$ and $160^\circ$ are most affected.

The time-of-flight (TOF) obtained from the radio frequency of the cyclotron 
(stop signal for TDCs) and 
the timing signal from each of the eight telescopes (start signal) is registered 
for each charged-particle event.
Typical count rates for target-in and target-out runs were 10 and 2 Hz, 
respectively. The dead time 
of the data acquisition system was typically $1-2$  \% and never exceeded 10  \%.


\section{Data reduction procedures and corrections}
\label{sec:Data reduction}


The $\Delta E - E$ technique is used to identify light charged particles ranging 
from protons to lithium 
ions. Good separation of all 
particles is obtained 
over their entire energy range and particle identification is straightforward. 

Energy calibration of all detectors is obtained from the data 
themselves~\cite{Tip04,Udo04}. Events in the 
$\Delta E - E$ bands are fitted with respect to the energy deposited in the 
two silicon detectors. This energy is determined from the detector thicknesses 
and tabulated energy loss 
values in silicon~\cite{Ziegler}. The $\Delta E_1$ detectors are further 
calibrated and checked using a 
5.48 MeV alpha source. For the energy calibration of the CsI(Tl) detectors, two 
parameterizations of the 
light output versus energy of the detected particle~\cite{Dan00,Tip04,Udo04} are used, one 
for hydrogen isotopes and 
another one for helium isotopes. Supplementary calibration points are provided 
by the H(n,p) reaction, as 
well as transitions to the ground state and low-lying states in the 
$^{12}$C(n,d)$^{11}$B, $^{16}$O(n,d)$^{15}$N and $^{28}$Si(n,d)$^{27}$Al 
reactions. The energy of each particle is obtained by adding 
the energy deposited in each element of the telescope.

Low-energy charged particles are stopped in the $\Delta E_1$ detector leading to 
a low-energy cutoff for 
particle identification of about 3 MeV for hydrogen isotopes and about 8 MeV for 
helium isotopes. The helium isotopes stopped in the $\Delta E_1$ 
detector are nevertheless 
analyzed, and a remarkably low cutoff, about 4 MeV, can be achieved for the 
experimental alpha-particle 
spectra. These alpha-particle events could obviously not be separated from 
$^3$He events in the same energy 
region, but the yield of $^3$He is about a factor of 30 smaller than the 
alpha-particle yield in the region of
8 MeV, where the particle identification works properly. The assumption that the relative yield 
of $^3$He is small is 
supported by the theoretical calculations in the evaporation peak region. 
In conclusion, the 
$^3$He yield is within the statistical uncertainties of the alpha-particle yield 
for alpha energies 
between 4 and 8 MeV. A consequence of this procedure is that the $^3$He spectra 
have a low-energy cutoff 
of about 8 MeV.

Knowing the energy calibration and the flight distances, the flight time for 
each charged particle from 
target to detector can be calculated and subtracted from the measured total 
TOF. The resulting neutron TOF is used for selection of charged-particle events 
induced by neutrons in the main peak of the incident neutron spectrum.

Background events, collected in target-out runs and analyzed in the same way as 
target-in events, are 
subtracted from the corresponding target-in runs, with SiO$_2$ and silicon targets,
after normalization to the same neutron fluence. 
 
Due to the finite target thickness, corrections for energy loss and particle loss
are applied to both targets individually. Details of the correction methods are
described in Refs.~\cite{Tip04,Pomp}. The cross sections for oxygen are obtained 
after subtraction of the silicon data from the SiO$_2$ data with proper 
normalization with respect to the number of silicon nuclei in the 
two targets.\footnote{
In the process of extracting the oxygen data, 
the silicon data of Ref.~\cite{Tip04} have been reanalyzed.
In doing so, we have adapted some changes and also found two mistakes.
See Ref.~\cite{Tip06} in this issue.
} 

Even if a great majority of the neutrons appears in the narrow full-energy peak at 95.6 
MeV, a significant fraction 
(about 13  \%) belongs to a tail extending towards lower energies, remaining 
after the TOF cut, see Fig.~\ref{fig:fig1}. The average neutron energy with 
the tail neutrons included is 94.0 MeV. The particle spectra have not been
unfolded with the neutron energy distribution, because it is anticipated that 
the energy variation of the cross sections is rather weak in the energy range of interest.
Furthermore, the data set is called 96 MeV (95.6 MeV) data, because the peak of the 
distribution is quite dominant and any structure observed at the high-energy
end of the ejectile spectra is due to the peak of the neutron energy dsitribution.
The $np$ cross section is, however, measured at the peak of the distribution
(95.6 MeV) and corrected for the tail contribution. The correction to 94.0 MeV
is performed using the known energy dependence of the $np$ cross section. 

Other corrections of the data are performed in analogy with the similar
experiment dealing with silicon and described in detail in the corresponding
publication~\cite{Tip04}. The data and method for the efficiency correction of the
CsI(Tl) detectors, reported in Ref.~\cite{Klu02} and used in Ref.~\cite{Tip04} and
the present work, have recently~\cite{Bli04} been corroborated by Monte Carlo
calculations. 

Absolute double-differential cross sections are obtained by normalizing the 
oxygen data to the 
number of recoil protons emerging from the CH$_2$ target. After selection of 
events in the main 
neutron peak and proper subtraction of the target-out and $^{12}$C(n,p) 
background contributions, the 
latter taken from a previous experiment, the cross section can be determined 
from the recoil proton 
peak, using $np$ scattering data~\cite{Rah01}. All data have been normalized 
using the $np$ scattering 
peak in the $20^\circ$ telescope.


\section{Theoretical models}
\label{sec:Theoretical models}


The present data have been compared with nuclear theory predictions, computed with the two 
nuclear reaction codes 
GNASH~\cite{Young,Cha97} and TALYS~\cite{Koning}. While GNASH has been widely 
used during the last years, 
TALYS is a new code that has just been released in the public domain. The GNASH calculation is 
performed at a neutron energy of 100 MeV with 
parameters given in a recent evaluation for medical purposes~\cite{ICRU} as described in Ref.~\cite{Tip04}. 
Since oxygen is at the boundary of the
mass range the TALYS code is aimed for, the code is described in some detail below.

Both GNASH and TALYS integrate direct, pre-equilibrium, and statistical nuclear 
reaction models into one 
calculation scheme and thereby give predictions for all the open reaction 
channels. Both codes use the 
Hauser-Feshbach model for sequential equilibrium decay and the exciton model for 
pre-equilibrium 
emission, though GNASH uses the one-component model, i.e. without isospin 
distinction of the excited nucleons, and TALYS uses the two-component model, 
see below. The angular distributions are obtained using the Kalbach 
systematics~\cite{Kal88}.

The purpose of TALYS is to simulate nuclear reactions that involve neutrons, 
photons, protons, deuterons, 
tritons, $^3$He and alpha particles in the 1 keV -- 200 MeV energy range. 
Predicted quantities include 
integrated, single- and double-differential cross sections, for both the 
continuum and discrete states, 
residue production and fission cross sections, gamma-ray production cross 
sections, etc. For the present 
work, single- and double-differential cross sections are of interest. To predict 
these, a calculation 
scheme is invoked which consists of a direct + pre-equilibrium reaction 
calculation followed by subsequent 
compound nucleus decay of all possible residual nuclides calculated by means of 
the Hauser-Feshbach model.

For the optical model potentials (OMP) of both neutrons 
and protons on $^{16}$O 
up to 200 MeV, the global OMP of 
Ref.~\cite{Kon03} was used. These potentials 
provide the necessary transmission coefficients for 
the statistical model calculations. 
Although the global neutron OMP has been validated for $A>24$, at the
high incident energy considered in this work, an adequate description of the basic scattering
observables is expected, at least for the incident neutron channel and the high 
energy inelastic scattering and charge-exchange leading to discrete states and 
the continuum. For the low-energy outgoing charged particles, the non-validated use 
of the global OMP may have larger consequences. Obviously, a system of a total 
of 17 nucleons can hardly be called statistical, and this short-coming may be 
reflected in the prediction of some of the observables that concern low 
emission energies.
For complex particles, the optical potentials were directly 
derived from the nucleon 
potentials using the folding approach of Watanabe~\cite{Wat58}. 
Finally, since applying the charged-particle OMP's for nuclides as light as 
$^{16}$O may be physically dubious, we renormalize the obtained OMP transmission coefficients
with the empirical non-elastic cross sections of Ref.~\cite{tri97}.

The high-energy end of the ejectile spectra are described by
pre-equili\-brium emission, which takes place after the first stage of the reaction but 
long before statistical 
equilibrium of the compound nucleus is attained. It is imagined that the 
incident particle step-by-step 
creates more complex states in the compound system and gradually loses its 
memory of the initial energy 
and direction. The default pre-equilibrium model of TALYS is the two-component 
exciton model~\cite{kon04,Kal86}. A remark similar to that given above for the 
OMP applies: the two-component exciton model for nucleon reactions has been 
tested, rather successfully, against basically all available
experimental nucleon spectra for $A>24$~\cite{kon04}. The current system 
$A=17$, falls outside that mass range, and does not entirely qualify as a 
system that can be handled by fully statistical models such as the exciton 
model.

We recall the basic formula of Ref.~\cite{kon04} for the exciton model cross section,
\begin{equation}
\frac{d\sigma_{k}^{EM}}{dE_{k}} = \sigma^{CF}
\sum_{p_{\pi}=p_{\pi}^{0}}^{p_{\pi}^{eq}}
\sum_{p_{\nu}=p_{\nu}^{0}}^{p_{\nu}^{eq}}
w_{k}(p_{\pi},h_{\pi},p_{\nu},h_{\nu},E_{k})
S_{pre}(p_{\pi},h_{\pi},p_{\nu},h_{\nu}),
\label{eq:pespec2}
\end{equation}
where $p_{\pi} (p_{\nu})$ is the proton (neutron) particle number and $h_{\pi} 
(h_{\nu})$ the proton (neutron) hole number, 
$\sigma^{CF}$ is the compound formation cross section, and $S_{pre}$ is the 
time-integrated strength which 
determines how long the system remains in a certain exciton 
configuration. The initial proton 
and neutron particle numbers are denoted $p_{\pi}^{0}=Z_{p}$ and 
$p_{\nu}^{0}=N_{p}$ with $Z_{p} (N_{p})$ 
being the proton (neutron) number of the projectile. In general, 
$h_{\pi}=p_{\pi}-p_{\pi}^{0}$ and $h_{\nu}=
p_{\nu}-p_{\nu}^{0}$, so that the initial hole numbers are zero, i.e. 
$h_{\pi}^{0}=h_{\nu}^{0}=0$, for primary 
pre-equilibrium emission. The pre-equilibrium part is calculated by 
Eq.~(\ref{eq:pespec2}), using $p_{\pi}^{eq}=p_{\nu}^{eq}=6$, 
whereas the remainder of the reaction flux is distributed through the 
Hauser-Feshbach model. In addition, the 
never-come-back approximation is adopted.

The emission rate $w_k$ for ejectile $k$ with spin $s_k$ is given by
\begin{equation}
w_{k}(p_{\pi},h_{\pi},p_{\nu},h_{\nu},E_{k}) = \frac{2s_{k}+1}{\pi ^{2}\hbar ^{3
}}\mu _{k} E_{k}
\sigma _{k,inv}(E_{k})
\frac{\omega (p_{\pi}-Z_{k},h_{\pi},p_{\nu}-N_{k},h_{\nu},E_{x})}
{\omega (p_{\pi},h_{\pi},p_{\nu},h_{\nu},E^{tot})},
\label{eq:emission2}
\end{equation}
where $\sigma _{k,inv}(E_{k})$ is the inverse reaction cross section as 
calculated from the optical model, 
and $\omega$ is the two-component particle-hole state density. 
The full reaction dynamics that leads to Eq.~(\ref{eq:pespec2}) is described in Refs.
\cite{kon04,Kal86}. We here restrict ourselves to the formulae given above since they contain
the model- and parameter-dependent quantities.
The expression 
for $S_{pre}$ contains the 
adjustable transition matrix element $M^{2}$ for each possible transition 
between neutron-proton exciton 
configurations. A proton-neutron ratio of 1.6 for the squared internal 
transition matrix elements was adopted 
to give the best overall agreement with experiment, i.e., 
$M_{\pi\nu}^{2}=M_{\nu\pi}^{2}=1.6M_{\pi\pi}^{2}=
1.6M_{\nu\nu}^{2}=1.6M^{2}$. 
For ${}^{16}$O, we use the following expression for the matrix element~\cite{kon04},
\begin{equation}
M^{2} =\frac{0.6}{A^{3}} \left[ 6.8 + \frac{4.2\times 10^{5}}
{(\frac{E^{\rm tot}}{n}+10.7)^{3}}\right], \label{matrix}
\end{equation}
where $n$ is the exciton number. Partial level density parameters $g_{\pi}=Z/17$ and 
$g_{\nu}=N/17$ were used in the 
equidistant spacing model for the partial level densities.
Finally, an effective surface interaction well depth $V=12$ MeV~\cite{kon04} was used.

At incident energies above several tens of MeV, the residual nuclides formed 
after binary emission may 
have so large excitation energy that the presence of additional fast particles 
inside the nucleus becomes 
possible. The latter can be imagined as strongly excited particle-hole pairs 
resulting from the first 
binary interaction with the projectile. The residual system is then clearly 
non-equilibrated and the 
excited particle that is high in the continuum may, in addition to the first 
emitted particle, also be 
emitted on a short time scale. This so-called multiple pre-equilibrium emission 
forms an alternative 
theoretical picture of the intra-nuclear cascade process, whereby the exact 
location and momentum of 
the particles are not followed, but instead the total energy of the system and 
the number of 
particle-hole excitations (exciton number).
In actual calculations, the particle-hole configuration of the residual nucleus 
after emission of the 
ejectile, is re-entered as initial condition in Eq.~(\ref{eq:pespec2}). When 
looping over all possible residual 
configurations, the multiple pre-equilibrium contribution is obtained. In TALYS, 
multiple pre-equilibrium 
emission is followed up to arbitrary order, though for 96 MeV only secondary 
pre-equilibrium emission is 
significant. 

It is well-known that semi-classical models, such as the exciton model, have 
always had some problems to 
describe angular distributions (essentially because it is based on a 
compound-like concept instead of a 
direct one). Therefore, as mentioned previously, the double-differential cross 
sections are obtained from 
the calculated energy spectra using the Kalbach systematics~\cite{Kal88}. 

To account for the evaporation peaks in the charged-particle spectra, multiple 
compound emission was 
treated with the Hauser-Feshbach model. In this scheme, all reaction chains are 
followed until all 
emission channels are closed. The Ignatyuk model~\cite{Ign75} has been adopted 
for the total level 
density to account for the damping of shell effects at high excitation energies.

For pre-equilibrium reactions involving deuterons, tritons, $^3$He and alpha 
particles, a statistical contribution 
from the exciton model is automatically calculated with the formalism described 
above. It is, however, 
well known that for nuclear reactions involving projectiles and ejectiles with 
different particle numbers, 
mechanisms like stripping, pick-up and knock-out play an important role and 
these direct-like reactions to the continuum
are not covered by the exciton model. Therefore, Kalbach has developed a 
phenomenological contribution for 
these mechanisms~\cite{Kal01}, which is included in TALYS. Among the advantages 
over the older method (which is included in GNASH) we mention here a better 
consideration of the available phase space through normalized particle-hole 
state densities and a better empirical determination of the pick-up, stripping, 
knock-out strength parameters, enabled by the more extensive experimental 
database that is now available. It has recently been 
shown (see Table I of Ref.~\cite{Ker02}) 
that for medium and heavy nuclides this method gives a considerable 
improvement over the older methods. The 
latter seemed to consistently 
underpredict neutron-induced reaction cross sections involving pick-up of one or 
a few nucleons.
In this paper, the two methods meet again, this time for the prediction of 
reactions on a light nucleus, and their performance will be compared in the
next section.

\section{Results and discussion}
\label{sec:Results and discussion}


\subsection{Experimental results}
\label{subsec:Experimental results}

Double-differential cross sections of $^{16}$O(n,xlcp) reactions, where lcp stands for 
light charged particle, at laboratory
angles of $20^\circ$, $40^\circ$, $100^\circ$ and 
$140^\circ$ for protons, deuterons, tritons, $^3$He and alpha particles are 
shown in Figs.$~\ref{fig:fig2} -~\ref{fig:fig6}$, respectively, all angles plotted
with the same cross section scale for each emitted particle to facilitate comparison of 
magnitudes. The choice of energy bin width depends on the energy 
resolution in the experiment, the thick target correction and 
acceptable statistics in each energy bin. The error bars 
in Figs.$~\ref{fig:fig2} -~\ref{fig:fig6}$ represent 
statistical uncertainties only.

The overall relative statistical uncertainties of individual points in the 
double-differential energy 
spectra at $20^\circ$ are typically 8 \% for protons, 13 \% for deuterons, 20 \% 
for tritons, 15 \% 
for $^3$He and 12 \% for alpha particles. As the angular distributions are 
forward-peaked, these values 
increase with angle. The systematic uncertainty contributions are due to thick 
target correction 
($1-20$ \%), collimator solid angle ($1-5$ \%), beam monitoring ($2-3$ \%), 
number of oxygen nuclei 
(0.1 \%), CsI(Tl) intrinsic efficiency (1 \%), particle identification (1 \%) and 
dead time ($<$\,0.1 \%). 
The uncertainty in the absolute cross section is about 5 \%, which is due to 
uncertainties in $np$ 
scattering angle, contribution from the low-energy continuum of the 
$^7$Li(p,n) spectrum to the $np$ 
scattering proton peak (3 \%), reference $np$ cross sections 
(2 \%)~\cite{Rah01}, statistics in the 
$np$ scattering proton peak (2 \%), carbon contribution (0.1 \%) and 
number of hydrogen nuclei (0.1 \%).

From Figs.$~\ref{fig:fig2} -~\ref{fig:fig6}$ it is obvious that the 
charged-particle emission at forward angles from 96 MeV 
neutron irradiation of oxygen is dominated by proton, deuteron and alpha 
particle channels. The yield of deuterons is about a factor of 3 lower than for protons 
and the spectra 
of the two other particle types studied in this work (tritons and $^3$He) are 
more than an order of 
magnitude weaker. All the spectra have more or less pronounced peaks at low 
energies (below $10-15$ MeV), 
the angular distributions of which are not too far from isotropy except for alpha 
particles, where the yield 
at backward angles is about four times weaker than at $20^\circ$. 
The low-energy peak is not fully observed in 
the $^3$He spectra due to the 8 MeV low-energy cutoff discussed in 
Sec.~\ref{sec:Data reduction}.

All the particle spectra at forward angles show relatively large yields at 
medium-to-high energies. 
The emission of high-energy particles is strongly forward-peaked and hardly 
visible in the backward 
hemisphere. It is a sign of particle emission before statistical equilibrium has 
been reached in the 
reaction process. In addition to this broad distribution of emitted particles, 
the deuteron spectra at 
forward angles show narrow peaks corresponding to transitions to the ground 
state and low-lying states 
in the final nucleus, $^{15}$N. These transitions are most likely due to 
pick-up of weakly bound protons 
in the target nucleus, $^{16}$O. A similar but less pronounced effect is observed 
in the proton spectra at forward angles. The structure observed in this case is
due to transitions to Gamow-Teller states and other low-lying states with considerable
single-particle strength~\cite{Nee82}. 

%
\subsection{Comparison with theoretical model calculations}
\label{subsec:Comparison with}

In Figs.$~\ref{fig:fig2} -~\ref{fig:fig6}$ the experimental results are presented 
together with theoretical 
model calculations.
The GNASH calculations of Ref.~\cite{ICRU} have been done for protons, deuterons 
and alpha particles,
whereas the TALYS calculations discussed in Sec.~\ref{sec:Theoretical models} 
have been performed for all five particle 
types. The TALYS calculations  
include a transformation of the calculated cross sections 
to the lab system. Also in the GNASH code a similar transformation from the c.m. 
to the lab system is performed using the kinematics of one-particle emission.
Differences between data given in the lab and c.m. systems are  
particularly significant in this case, because oxygen is such a light nucleus.

Fig.~\ref{fig:fig2} shows the comparison between the double-differential (n,px) 
experimental spectra and the 
calculations based on the TALYS and GNASH models. For protons above 25 MeV, both 
calculations give a reasonably good description of the spectra, although the 
calculated $20^\circ$ cross sections, in particular the TALYS ones, fall below the 
experimental data. The low-energy statistical peak below 15 MeV in the spectra is 
considerably overpredicted by the two codes. The overestimate is particularly strong
at backward angles for TALYS and at forward angles for GNASH.

The situation is quite different for the deuteron spectra (Fig.~\ref{fig:fig3}). 
None of the calculations do account very well for the data, although the GNASH code
gives a reasonable description of the angular dependence of the cross section. For the TALYS code 
deviations between data and calculations of a factor of two or more are present. At forward angles 
the high-energy part is strongly overestimated, in particular by the TALYS code, 
indicating problems in the hole-strength treatment. It is obvious, however, that
efforts have been spent in these calculations to include individual hole-state 
strengths. Such strengths are not included in the GNASH calculations, but
in spite of this the average behavior of the cross section at high energies is
in fair agreement with the data. Like for the proton spectra, the statistical peak 
is overpredicted by the TALYS calculations essentially at all angles, whereas the
GNASH calculations seem to do a slightly better job in this case. 

For tritons (Fig.~\ref{fig:fig4}), the TALYS calculation gives a fairly good  
description of the experimental data, except that it fails to account for an
intensity bump around 15 MeV observed at forward angles. 

The general trends of the forward-angle $^3$He data (Fig.~\ref{fig:fig5}) 
are reasonably well described in the TALYS calculations although the cross sections 
are underestimated by a large factor. At backward angles the yield is
very small and it is difficult to make quantitative comparisons.

The overall shapes of the alpha particle spectra (Fig.~\ref{fig:fig6}) are reasonably
well described by the two models. The GNASH calculations, however, overpredict 
the cross sections at forward angles and underpredict them at large angles, 
whereas the TALYS calculations do the opposite, i.e., underpredict at small
angles and overpredict at large angles.

The ability of the models to account for the low-energy peak caused by 
evaporation processes (and for $\alpha$ particles also 3$\alpha$ breakup of 
$^{12}$C) is not 
impressive. In general, the models tend to overpredict the cross sections. It 
should, however, be kept in 
mind that the peak maximum is close to (for $^3$He below) the low-energy cutoff, 
which complicates the 
comparison. Another complication in this context is that the 
GNASH cross sections although
given in the lab system, are calculated using the kinematics of 
one-particle emission~\cite{Young,Cha97} for the c.m.-to-lab transformation, 
which obviously is an approximation.

Experimental angular distributions at low, medium and high ejectile energies 
are shown in Figs.$~\ref{fig:fig7} -~\ref{fig:fig11}$ for protons, deuterons,
tritons, $^3$He and alpha particles, respectively. The angular distributions are
fitted by a simple two parameter function, $a\exp(b\cos\theta)$~\cite{Kal88}.
The data are compared with angular distributions calculated on the basis of the
GNASH and TALYS models. In general, the TALYS model gives a weaker angular
dependence than the data, whereas the GNASH model, although being closer to the
data, tends to give a slightly steeper angular variation.

A conspicuous deviation from the experimental angular distribution is 
seen for the TALYS prediction at the lowest outgoing energies, e. g. at $8-12$ 
MeV in Fig.~\ref{fig:fig7}. We think this is attributed to wrong partial spectrum
contributions to the total spectrum. The slightly forward-peaked angular 
distribution suggests that the spectrum at these emission energies is not as 
compound-dominated as the TALYS calculation suggests. Instead secondary, and even 
tertiary, pre-equilibrium emission may not be negligible even in the evaporation
peak. Multiple pre-equilibrium emission is taken into account in TALYS but only 
contributes at somewhat higher emission energies. A way to make multiple 
pre-equilibrium (processes) relatively more important is to reduce the compound
nucleus emission contribution, but we find that the predicted evaporation peak is 
rather insensitive to parameter variations. Hence, this is an open problem for 
TALYS, which apparently has been solved for the GNASH calculation.

\subsection{Integrated spectra}
\label{subsec:Integrated spectra}

For each energy bin of the light-ion spectra, the experimental angular 
distribution is fitted by a simple 
two-parameter function, $a\exp(b\cos\theta)$~\cite{Kal88}, as exemplified in
the previous subsection (Figs.$~\ref{fig:fig7} -~\ref{fig:fig11}$).
This allows extrapolation of double-differential cross sections to very forward 
and very backward angles. 
In this way coverage of the full angular range is obtained. By integration of 
the angular distribution, 
energy-differential cross sections ($d\sigma /dE$) are obtained for each 
ejectile. These are shown in Fig.~\ref{fig:fig12}
together with theoretical calculations. For all ejectiles 
both calculations give a fair description of the energy dependence. Both calculations 
are in good agreement with 
the proton experimental data over the whole energy range, although the 
calculation for (n,p) reactions to discrete states is underestimating the 
data. A study of the spectroscopic strengths for these 
states would be welcome. 
Concerning the deuteron
spectra, the GNASH calculations are in good agreement with the data, whereas the TALYS code
gives cross sections a factor of two or more larger than the experimental ones 
at energies above 30 MeV. In the case of 
alpha particles, the GNASH calculation tends to
overpredict the high-energy part of the spectrum, and the TALYS calculations fall 
below the data above an alpha particle energy of 25 MeV. The energy dependence of the
triton and $^3$He spectra are well described by the TALYS code, but in both cases 
the calculation falls below the data above about 20 MeV.

The production cross sections are deduced by integration of the 
energy-differential spectra (see Table~\ref{tab:tab1}). 
To be compared with the calculated cross sections, the experimental values 
in Table~\ref{tab:tab1} have to be corrected for the undetected 
particles below the low-energy cutoff. This is particularly important for $^3$He 
because of the high cutoff energy. The corrections obtained with TALYS 
seem to be too small in some cases,
in particular for the (n,x$\alpha$) production cross section. 
This is illustrated in Fig.~\ref{fig:fig12}, bottom panel, where the TALYS curve falls well
below the experimental $d\sigma /dE$ data in the 4--7 MeV region.

The proton, deuteron, triton, and alpha particle production cross sections are 
compared with previous data at lower energies~\cite{Ben98c} in Fig.~\ref{fig:fig13}. 
There seems to be general agreement between the trends of the 
previous data and the present data points. The curves in this figure are based 
on a GNASH calculation~\cite{ICRU}.   


\section{Conclusions and outlook}
\label{sec:Conclusions and outlook}


In the present paper, we report an experimental data set on light-ion production 
in oxygen induced by 96 MeV 
neutrons. Experimental double-differential cross sections 
($d^2\sigma/d\Omega dE$) are 
measured at eight angles between $20^\circ$ and $160^\circ$. Energy-differential 
($d\sigma /dE$) and 
production cross sections are obtained for the five types of outgoing particles. 
Theoretical calculations 
based on nuclear reaction codes including direct, pre-equilibrium and 
statistical models give 
generally a good account of the magnitude of the experimental cross sections. 
For proton emission, the 
shape of the spectra for the double-differential and energy-differential cross 
sections are well described. 
The calculated and the experimental alpha-particle spectra are also in fair 
agreement with the exception 
of the high energy part, where the GNASH model predicts higher yield and the TALYS model 
lower yield than 
experimentally observed. For the proton evaporation peak, the global TALYS
calculation overestimates the data. A future activity should be an adjustment 
of the responsible OMP and level density parameters (as was done in the case of 
GNASH) instead of relying on a full global prediction.
For the 
other complex ejectiles (deuteron, triton and $^3$He) there are important 
differences between theory and 
experiment in what concerns the shape of the spectra at various angles. 
We think this is due to the use of statistical models such as the 
Hauser-Feshbach model and the pre-equilibrium exciton model in mass ranges 
where these models become suspect, and the absence of a break-up model in the 
theoretical analysis. Apart from the aforementioned break-up model, predictions
of emission of alpha particles may be particularly sensitive to a correct
knock-out model and the use of adequate complex particle optical model
potentials. Stripping and knock-out models, level densities, optical model
and omission of break-up reactions may all add up to problems for something as light
as oxygen. This needs to be studied in much more detail. Finally, the magnitude
of the angle-integrated cross sections is reasonably well accounted for.

For the further development of the field, data at even higher energies are 
requested. The results 
suggest that the MEDLEY facility, which was used in the present work, should be 
upgraded to work also 
at 180 MeV, i.e., the maximum energy of the TSL neutron beam facility. At 
present, a new neutron beam 
facility is under commissioning at TSL~\cite{Pom04}, covering the same energy range, but with 
a projected intensity 
increase of a factor five. This will facilitate measurements at energies higher
than in the present work.

\section*{Acknowledgments}

This work was supported by the Swedish Natural Science Research Council, the 
Swedish Nuclear Fuel and 
Waste Management Company, the Swedish Nuclear Power Inspectorate, Ringhals AB, 
the Swedish Defence 
Research Agency, and the Swedish International Development Authority. 
The authors wish to thank the The Svedberg Laboratory for 
excellent support. One of the 
authors (U.T.) wishes to express his gratitude to the Thai Ministry of 
University Affairs and to the 
International Program in the Physical Sciences at Uppsala University.


\pagebreak

\section*{Figure captions}

\begin{enumerate}



\item
\label{fig:fig1}
Neutron energy distribution with the time-of-flight criterion applied derived 
from $np$ scattering data at an angle of 20 degrees. The peak (95.6 MeV), 
median (95.1 MeV) and average (94.0 MeV) are indicated by solid, dashed and 
dotted vertical lines, respectively.


\item
\label{fig:fig2}
Experimental double-differential cross sections (filled circles) of the O(n,px) 
reaction at 96 MeV at four 
laboratory angles. The curves indicate theoretical calculations based on 
GNASH (dotted) and TALYS (solid).


\item
\label{fig:fig3}
Experimental double-differential cross sections (filled circles) of the O(n,dx) 
reaction at 96 MeV at four 
laboratory angles. The curves indicate theoretical calculations based on 
GNASH (dotted) and TALYS (solid).


\item
\label{fig:fig4}
Experimental double-differential cross sections (filled circles) of the O(n,tx) 
reaction at 96 MeV at four laboratory angles. 
The curve indicates theoretical calculations based on TALYS.


\item
\label{fig:fig5}
Experimental double-differential cross sections (filled circles) of the 
O(n,$^3$Hex) reaction at 96 MeV at four laboratory angles. 
The curve indicates theoretical calculations based on TALYS.


\item
\label{fig:fig6}
Experimental double-differential cross sections (filled circles) of the 
O(n,$\alpha$x) reaction at 96 MeV at four laboratory angles. 
The curves indicate theoretical calculations based on 
GNASH (dotted) and TALYS (solid).


\item
\label{fig:fig7}
Angular distributions of O(n,px) cross section at ejectile energies of 
8 - 12 MeV (filled circles), 40 - 44 MeV (filled triangles) and 68 - 72 MeV
(open squares). The dashed curves are fits to the data and the dotted and 
solid curves represent calculations based on the GNASH and TALYS models,
respectively.


\item
\label{fig:fig8}
Same as Fig.~\ref{fig:fig7} but for the O(n,dx) cross section.


\item
\label{fig:fig9}
Same as Fig.~\ref{fig:fig7} but for the O(n,tx) cross section. No calculations
based on the GNASH model are available for tritons.

\item
\label{fig:fig10}
Angular distributions of O(n,$^3$Hex) cross section at ejectile energies of 
10 - 15 MeV (filled circles), 40 - 45 MeV (filled triangles) and 65 - 70 MeV
(open squares). The dashed curves are fits to the data and the solid 
curves represent calculations based on the TALYS model.


\item
\label{fig:fig11}
Same as Fig.~\ref{fig:fig10} but for the O(n,$\alpha$x) cross section. The
dotted curves represent calculations based on the GNASH model.

\item
\label{fig:fig12}
Experimental energy-differential cross sections (filled circles) for 
neutron-induced p, d, t, $^3$He and 
$\alpha$ production at 96 MeV. The curves indicate theoretical calculations 
based on GNASH (dotted) and TALYS (solid).


\item
\label{fig:fig13}
Neutron-induced a) proton, b) deuteron, c) triton, and d) alpha particle 
production cross section as a function of neutron energy. 
The full circles are from 
the present work, whereas the open circles are from previous work~\cite{Ben98c}. 
The curves are based on a GNASH calculation. 
The data as well as the calculations correspond to cutoff energies of 
6 MeV for protons and deuterons and 12 MeV for tritons and alpha particles.
Note that the cutoff energies are different from those in 
Table ~\ref{tab:tab1}.

\end{enumerate}

\pagebreak

\begin{table}
\caption{Experimental production cross sections for protons,
deuterons, tritons, $^3$He and alpha particles from the present
work. The experimental data in the second column have been
obtained with cutoff energies of 2.5, 3.0, 3.5, 8.0 and 4.0 MeV
for p, d, t, $^3$He and alpha particles, respectively. The third
and forth columns show data corrected for these cutoffs, using the
GNASH (Ref.~\protect\cite{ICRU}) and the TALYS calculations of the
present work, respectively. Theoretical values resulting from GNASH and TALYS
calculations are in the fifth and sixth columns, respectively.}
\label{tab:tab1}
\begin{tabular}{cccccc}
\hline \hline \multicolumn{1}{c}{$\sigma_{prod}$} &
\multicolumn{1}{c}{Experiment} & \multicolumn{2}{c}{Experiment
[cutoff corr.]} &
\multicolumn{2}{c}{Theoretical calculation}\\
& (mb)&   GNASH & TALYS & GNASH & TALYS\\
\hline
(n,px)        & 224$\pm$11  & 248 & 231 & 259.9 & 221.7 \\
(n,dx)        &  72$\pm$4   & 80  & 73  & 73.4  & 131.3 \\
(n,tx)        &  20$\pm$1   & --  & 20  & --    & 10.6  \\
(n,$^3$Hex)   & 6.9$\pm$0.6 & --  & 8.7 & --    & 8.2   \\
(n,$\alpha$x) & 132$\pm$7   & 218 & 132 & 224.7 & 88.4  \\
\hline
\end{tabular}
\end{table}



\begin{thebibliography}{99}

%
\bibitem{Nee82}
G. A. Needham,  F. P. Brady, D. H. Fitzgerald, J. L. Romero, J. L. Ullmann, 
J. W. Watson, C. I. Zanelli, N. S. P. King, and G. R. Satchler, 
Nucl. Phys. {\bf A385}, 349 (1982).

\bibitem{Sub86}
T. S. Subramanian, J. L. Romero, F. P. Brady, D. H. Fitzgerald, R. Garrett,
G. A. Needham, J. L. Ullmann, J. W. Watson, C. I. Zanelli, D. J. Brenner, and 
R. E. Prael, Phys. Rev. C {\bf 34}, 1580 (1986).

\bibitem{Ben98a}
S. Benck, I. Slypen, J.P. Meulders, and V. Corcalciuc, Eur. Phys. J. {\bf 
3}, 149 (1998).

\bibitem{Ben98b}
S. Benck, I. Slypen, J.P. Meulders, and V. Corcalciuc, Eur. Phys. J. {\bf 
3}, 159 (1998).

\bibitem{Ben98c}
S. Benck, I. Slypen, J.P. Meulders, and V. Corcalciuc, Phys. Med. Biol. {\bf 
43}, 3427 (1998).

\bibitem{Ore98}
R. Orecchia, A. Zurlo, A. Loasses, M. Krengli, G. Tosi, S. Zurrida, P. Zucali, 
and U. Veronesi, Eur. J. Cancer 
{\bf 34}, 459 (1998).

\bibitem{Sch01}
D.L. Schwartz, J. Einck, J. Bellon, and G.E. Laramore, Int. J. Radiat. Oncol. 
Biol. Phys. {\bf 50}, 449 (2001).

\bibitem{Lar95}
G.E. Laramore and T.W. Griffin, Int. J. Radiat. Oncol. Biol. Phys. {\bf 32}, 879 
(1995).

\bibitem{Bar00} 
D.T. Bartlett, R. Grillmaier, W. Heinrich, L. Lindborg, D. O'Sullivan, 
H. Schraube, M. Silari, and L. Tommasino,
Radiat. Res. Congress Proceedings {\bf 2}, 719 (2000).  
See also the Proceedings of the International Conference on
Cosmic Radiation and Aircrew Exposure, Radiat. Prot. Dosim. {\bf 86(4)} (1999).
 
\bibitem{Single}
Single-Event Upsets in Microelectronics, topical issue, eds. H.H.K. Tang and N. 
Olsson, Mat. Res. Soc. 
Bull. {\bf 28} (2003).

\bibitem{Cha99}
M.B. Chadwick and E. Normand, IEEE Trans. Nucl. Sci. {\bf 46}, 1386 (1999).

\bibitem{High}
High and Intermediate energy Nuclear Data for Accelerator-driven Systems 
(HINDAS), European contract 
FIKW-CT-2000-00031, coord. J.P. Meulders.

\bibitem{Hind}
A. Koning, H. Beijers, J. Benlliure, O. Bersillon, J. Blomgren, J. Cugnon, M. 
Duijvestijn, Ph. Eudes, D. Filges, 
F. Haddad, S. Hilaire, C. Lebrun, F.-R. Lecolley, S. Leray, J.-P. Meulders, R. 
Michel, R.-D. Neef, R. Nolte, 
N. Olsson, E. Ostendorf, E. Ramstr\"{o}m, K.-H. Schmidt, H. Schuhmacher, I. 
Slypen, H.-A. Synal, and R. Weinreich, 
J. Nucl. Sci. Tech., Suppl. {\bf 2}, 1161 (2002).

\bibitem{Cec79}
R.A. Cecil, B.D. Anderson, and R. Madey, Nucl. Instr. Meth. {\bf 161}, 439 
(1979).

\bibitem{Blo03}
J. Blomgren and N. Olsson,
Radiat. Prot. Dosim. {\bf 103(4)}, 293 (2003).

\bibitem{Dan00} 
S. Dangtip, A. Ata\c{c}, B. Bergenwall, J. Blomgren, K. Elmgren, C. Johansson, 
J. Klug, N. Olsson, G. Alm Carlsson, 
J. S{\"o}derberg, O. Jonsson, L. Nilsson, P.-U. Renberg, P. Nadel-Turonski, C. 
Le Brun, F.R. Lecolley, J.F. Lecolley, 
C. Varignon, Ph. Eudes, F. Haddad, M. Kerveno, T. Kirchner, and C. Lebrun, Nucl. 
Instr. Meth. Phys. Res. A {\bf452}, 
484 (2000).

\bibitem{Tip04} 
U. Tippawan, S. Pomp, A. Ata\c{c}, B. Bergenwall, J. Blomgren, S. Dangtip, 
A. Hildebrand, C. Johansson, J. Klug, P. Mermod,  L. Nilsson, M. \"{O}sterlund,
N. Olsson, K. Elmgren, O. Jonsson, A. V. Prokofiev, P.-U. Renberg, P. 
Nadel-Turonski, V. Corcalciuc, Y. Watanabe, and A. J. Koning, Phys. Rev. C {\bf69}, 
064609 (2004). 

\bibitem{Udo04}
U. Tippawan, Doctoral thesis, Chiang Mai University (2004) (unpublished).

\bibitem{Klu02} 
J. Klug, J. Blomgren, A. Ata\c{c}, B. Bergenwall, S. Dangtip, K. Elmgren, C. 
Johansson, N. Olsson, S. Pomp, 
A.V. Prokofiev, J. Rahm, U. Tippawan, O. Jonsson, L. Nilsson, P.-U. Renberg, P. 
Nadel-Turonski, A. Ringbom, 
A. Oberstedt, F. Tovesson, V. Blideanu, C. Le Brun, J.F. Lecolley, F.R. 
Lecolley, M. Louvel, N. Marie, 
C. Schweitzer, C. Varignon, Ph. Eudes, F. Haddad, M. Kerveno, T. Kirchner, C. 
Lebrun, L. Stuttg{\'e}, I. Slypen, 
A. Smirnov, R. Michel, S. Neumann, and U. Herpers, Nucl. Instr. Meth. Phys. Res. 
A {\bf489}, 282 (2002).

\bibitem{Smi95} 
A.N. Smirnov, V.P. Eismont, and A.V. Prokofiev, Rad. Meas. {\bf 25}, 151 (1995).

\bibitem{Rah01} 
J. Rahm, J. Blomgren, H. Cond\'{e}, S. Dangtip, K. Elmgren, N. Olsson, T. 
R\"{o}nnqvist, R. Zorro, O. Jonsson, 
L. Nilsson, P.-U. Renberg, A. Ringbom, G. Tibell, S.Y. van der Werf, T.E.O. 
Ericson, and B. Loiseau, Phys. Rev. C 
{\bf 63}, 044001 (2001).

\bibitem{Ziegler} 
J.F. Ziegler, TRIM Code, version 95.4, IBM-Research, 1995.

%
\bibitem{Pomp}
S. Pomp, Internal note (unpublished).

\bibitem{Tip06}
U. Tippawan, S. Pomp, A. Ata\c{c}, B. Bergenwall, J. Blomgren,
S. Dangtip, A. Hildebrand, C. Johansson, J. Klug, P. Mermod,
L. Nilsson, M. \"Osterlund, N. Olsson, K. Elmgren, O. Jonsson,
A.V. Prokofiev, P.-U. Renberg, P. Nadel-Turonski, V. Corcalciuc,
Y. Watanabe, A. Koning,
Phys. Rev. C {\bf XX}, xxxxxx(E) (2006).

\bibitem{Bli04}
V. Blideanu, F.R. Lecolley, J.F. Lecolley, T. Lefort, N. Marie, A. Ata\c{c}, G. Ban, 
B. Bergenwall, J. Blomgren, S. Dangtip, K. Elmgren, Ph. Eudes, Y. Foucher, A. Guertin, 
F. Haddad, A. Hildebrand, C. Johansson, O. Jonsson, M. Kerveno, T. Kirchner, J. Klug, 
Ch. Le Brun, C. Lebrun, M. Louvel, P. Nadel-Turonski, L. Nilsson, N. Olsson, S. Pomp, 
A.V. Prokofiev, P.-U. Renberg, G. Rivi{\`e}re, I. Slypen, L. Stuttg{\'e}, U. Tippawan, and 
M. \"{O}sterlund, Phys. Rev. C {\bf 70}, 014607 (2004).


%
%
%
\bibitem{Young}
P.G. Young, E.D. Arthur, and M.B. Chadwick,  Los Alamos National Laboratory 
Report No LA-12343-MS 
(1992), GNASH-FKK version gn9cp0, PSR-0125.

\bibitem{Cha97}
M.B. Chadwick, P.G. Young, R.E. MacFarlane, and A.J. Koning, Proc. of 2nd Int. 
Conf. Accelerator-Driven 
Transmutation Technologies and Applications, Kalmar, Sweden, 1996, edited by H. 
Cond{\'e} (Gotab, Stockholm, 
Sweden, 1997), p. 483.

\bibitem{Koning}
A.J. Koning, S. Hilaire, and M.C. Duijvestijn, TALYS-0.64 User Manual, December 5, 2004,
NRG Report 21297/04.62741/P FAI/AK/AK.

\bibitem{ICRU}
ICRU Report {\bf 63}, International Commission on Radiation Units and 
Measurements, Bethesda, MD, March 2000.
 
\bibitem{Kal88}
C. Kalbach, Phys. Rev. C {\bf 37}, 2350 (1988).

\bibitem{Kon03}
A.J. Koning and J.P. Delaroche, Nucl. Phys. A {\bf 713}, 231 (2003).

\bibitem{Wat58}
S. Watanabe, Nucl. Phys. {\bf 8}, 484 (1958).

\bibitem{tri97} 
R.K. Tripathi, F.A. Cucinotta, and J.W. Wilson, Nucl. Instr. Meth. Phys. Res. B {\bf 117},
347 (1996).

R.K. Tripathi, F.A. Cucinotta, and J.W. Wilson, NASA technical
paper 3621, January 1997.

\bibitem{kon04} A.J. Koning, and M.C. Duijvestijn, Nucl. Phys. {\bf A744}, 15 
(2004).

\bibitem{Kal86}
C. Kalbach, Phys. Rev. C {\bf 33} , 818 (1986).

%
%
\bibitem{Ign75}
A.V. Ignatyuk, G.N. Smirenkin, and A.S. Tishin, Sov. J. Nucl. Phys. {\bf 21}, 
255 (1975).

\bibitem{Kal01}
C. Kalbach, Users manual for PRECO-2000, Duke University (2001).

C. Kalbach, Phys. Rev. C {\bf 71}, 034606 (2005).

\bibitem{Ker02}
M. Kerveno, F. Haddad, Ph. Eudes, T. Kirchner, C. Lebrun, I. Slypen, J. P. 
Meulders, C. Le Brun, F. R. Lecolley, 
J. F. Lecolley, M. Louvel, F. Lef{\`e}bvres, S. Hilaire, and A. J. Koning, Phys. 
Rev. C {\bf 66}, 014601 (2002).

%
%
%
%
%
%
\bibitem{Pom04}
S.~Pomp, A.V.~Prokofiev, J. Blomgren, C. Ekstr{\"o}m, O.~Jonsson, 
D. Reistad, V. Ziemann, N. Haag, A.~Hildebrand, L. Nilsson, B.~Bergenwall,
C.~Johansson, P.~Mermod, N.~Olsson, M.~{\"O}sterlund, and U.~Tippawan, 
The new Uppsala neutron beam facility, 
Proceedings of International Conference on 
Nuclear Data for Science and Technology, 
Santa Fé, NM, September 26 - October 1 2004, 
AIP Conference Proceedings No. 769 (Melville, New York, 2005), 780.
\end{thebibliography}
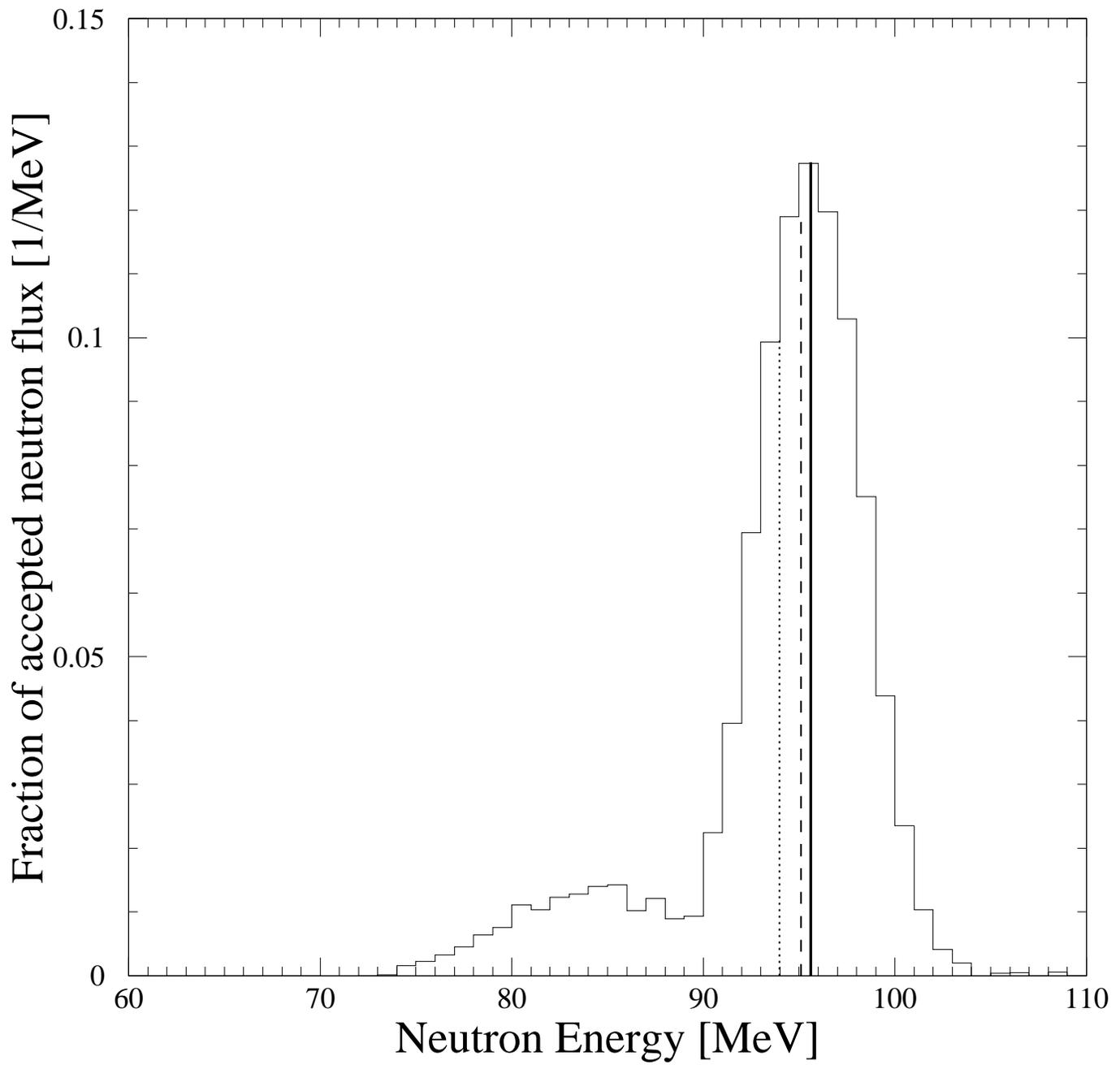
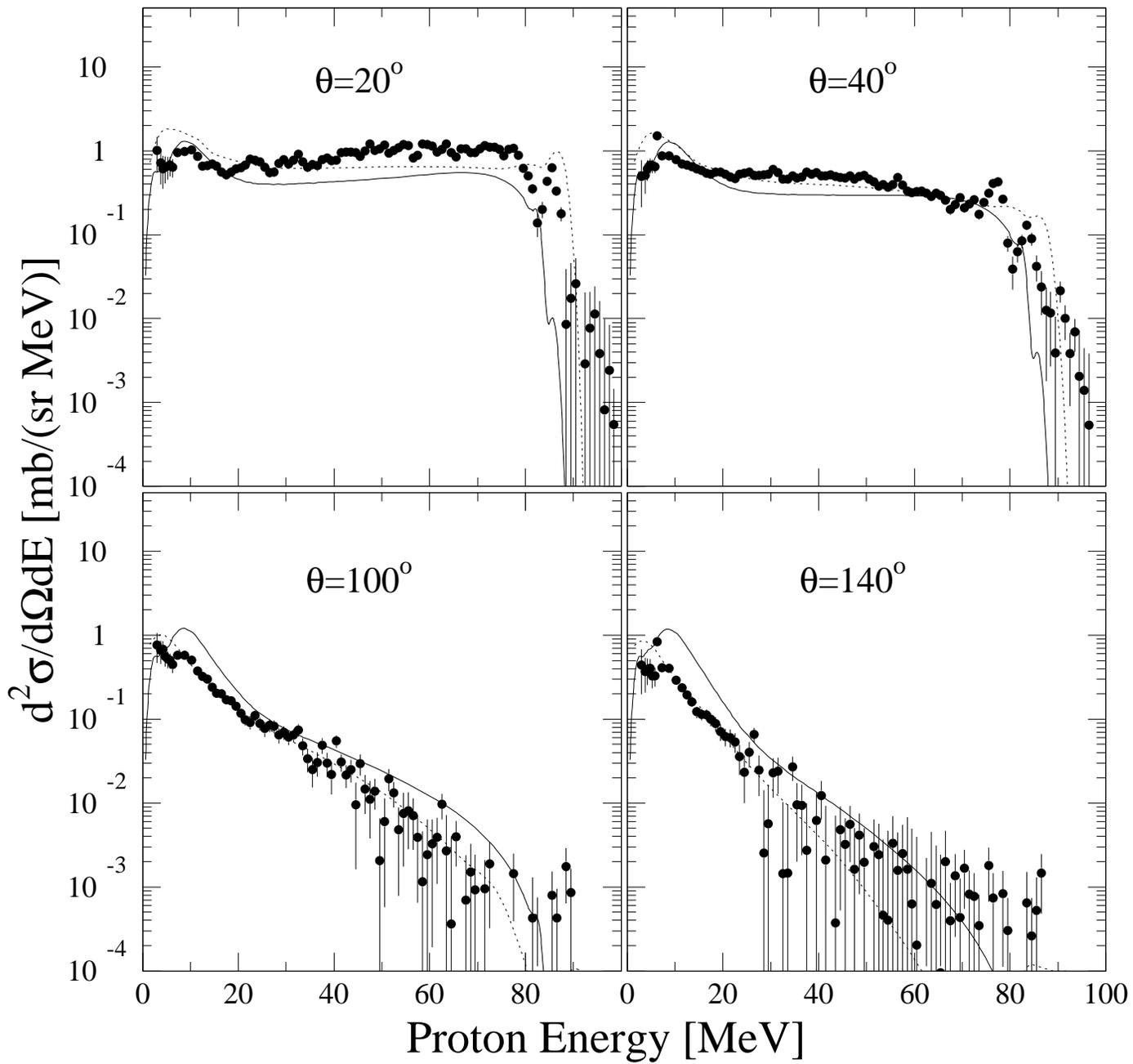
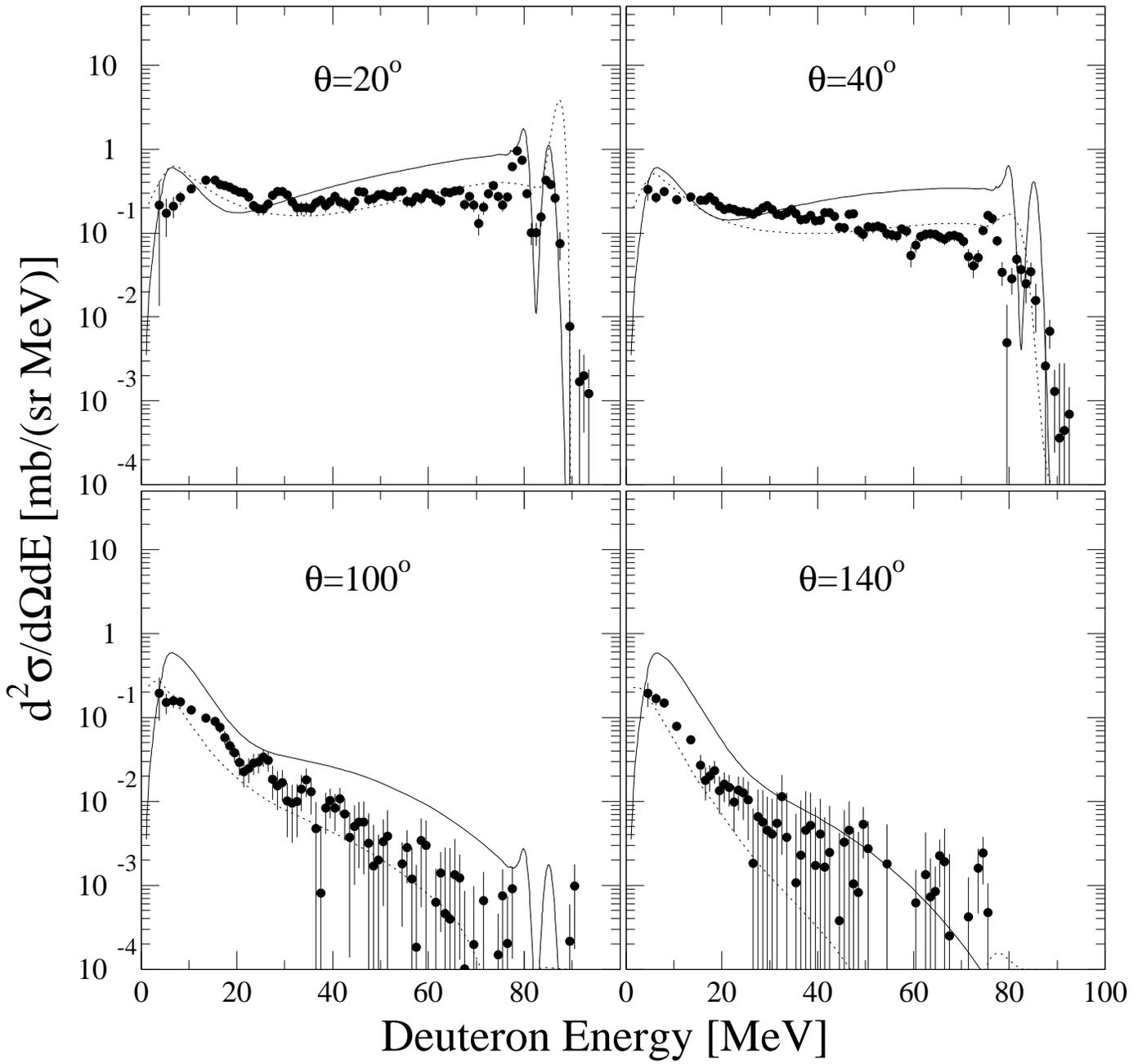
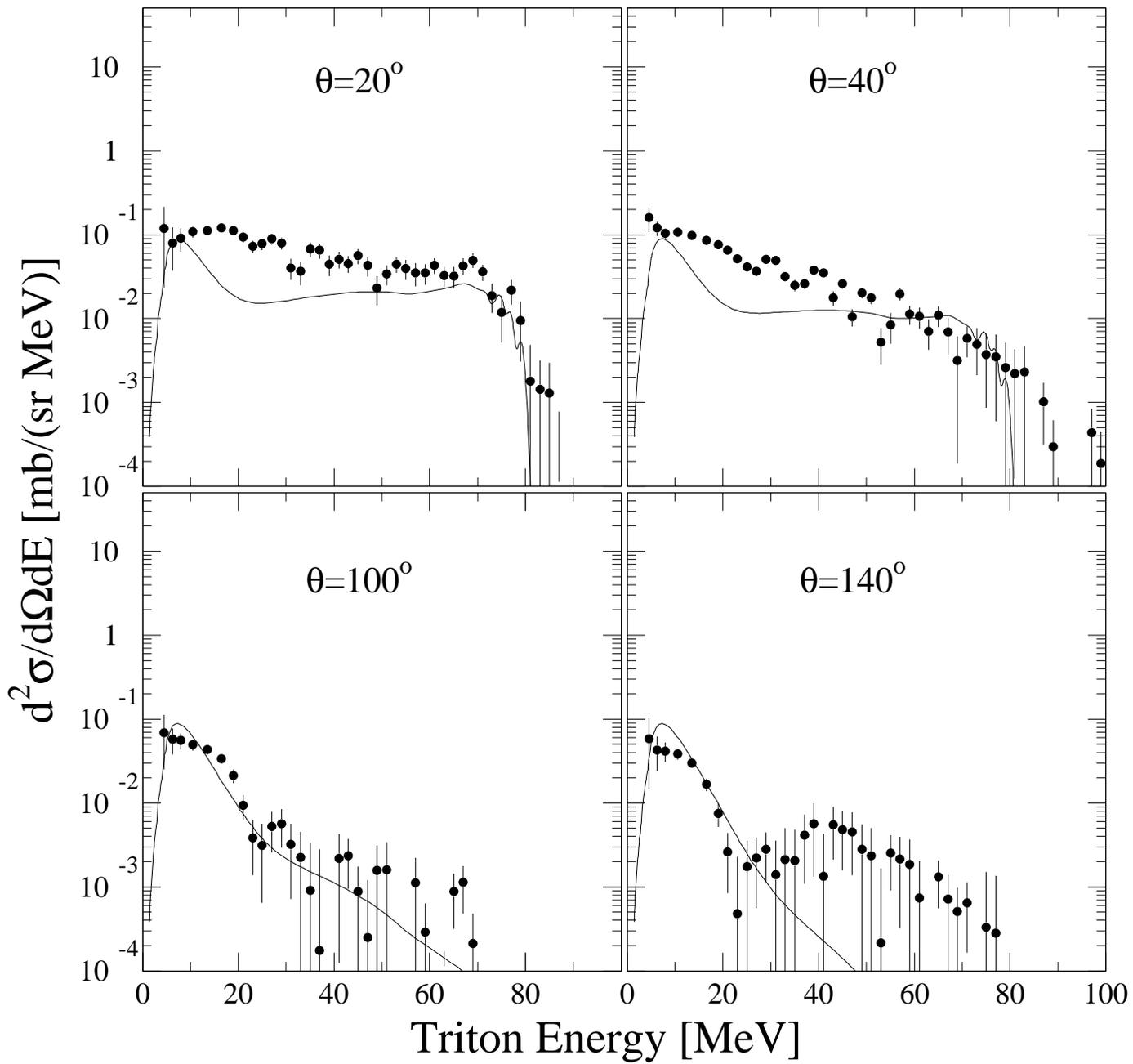
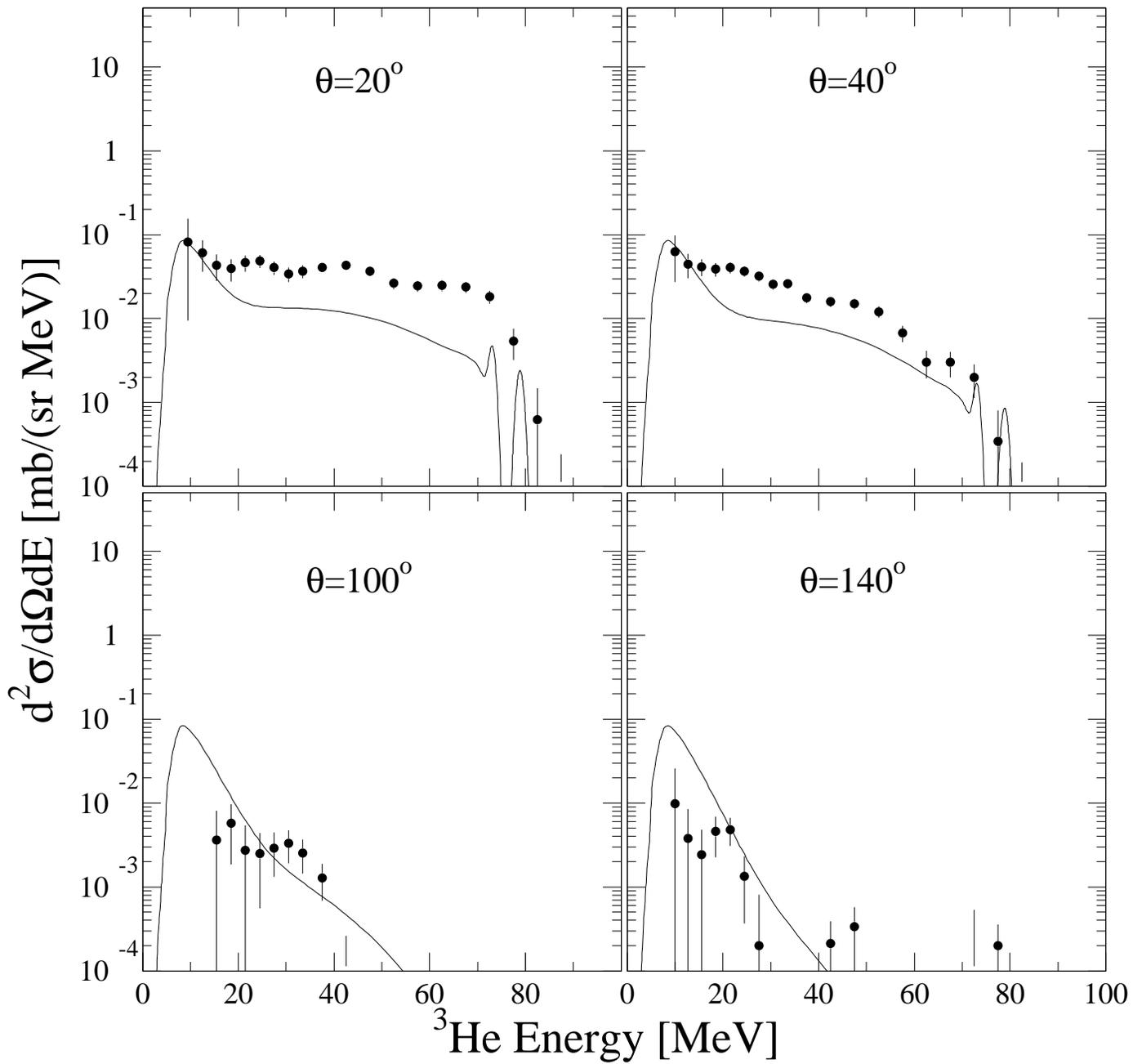
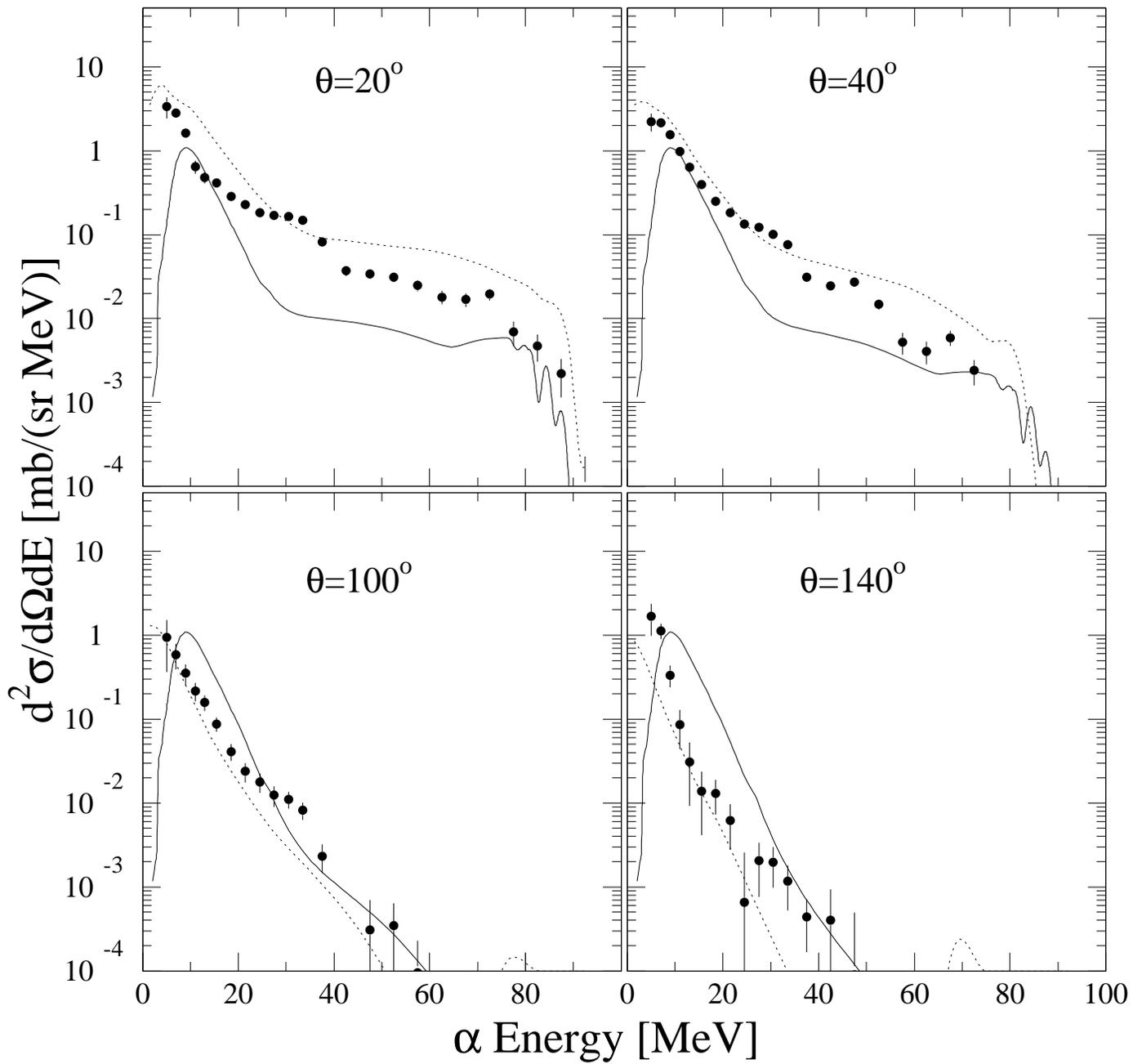
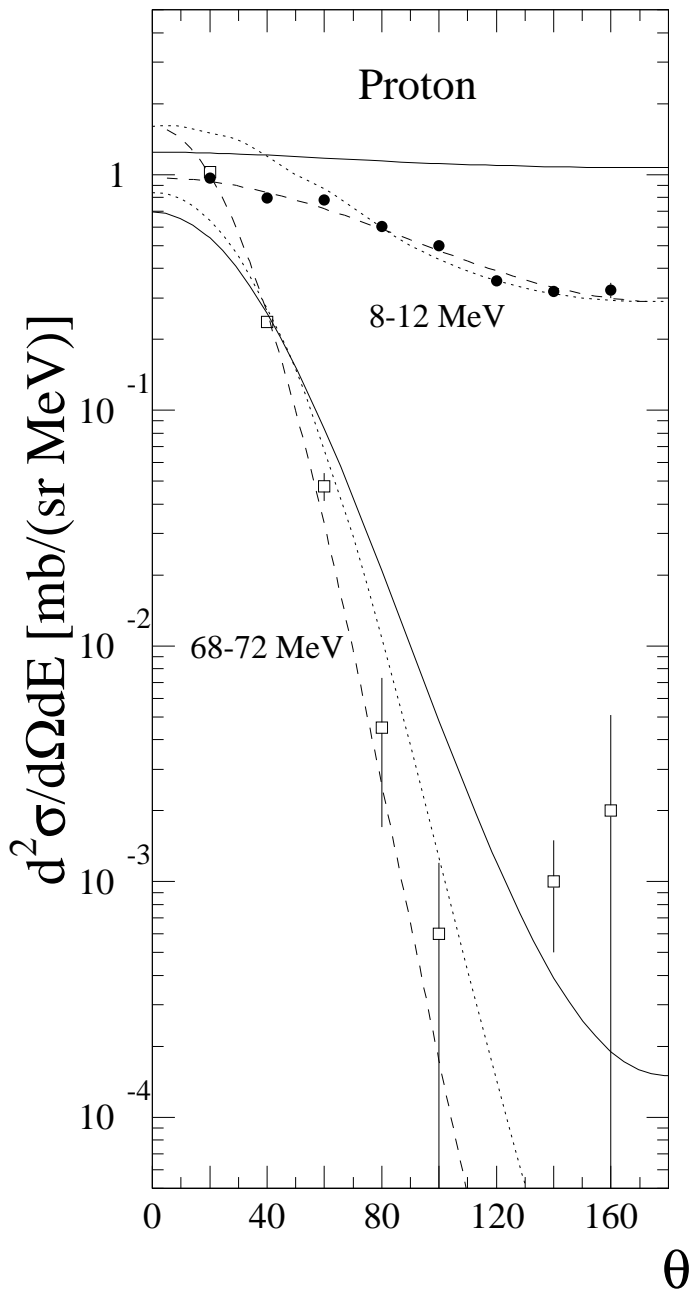
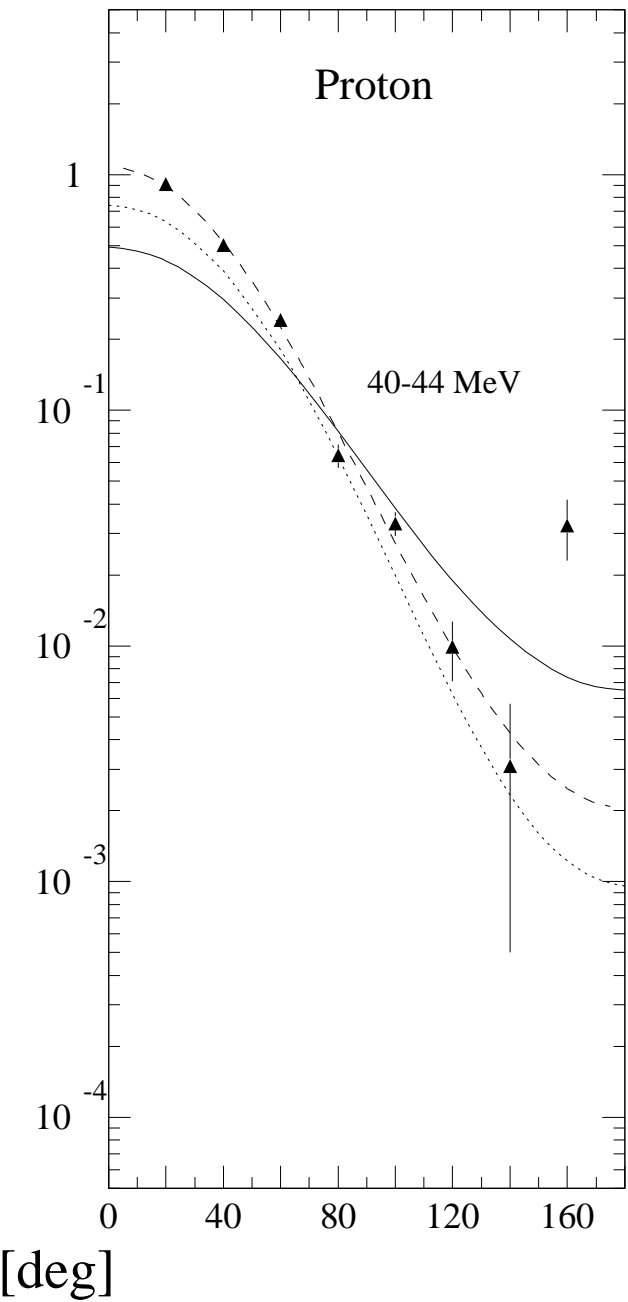
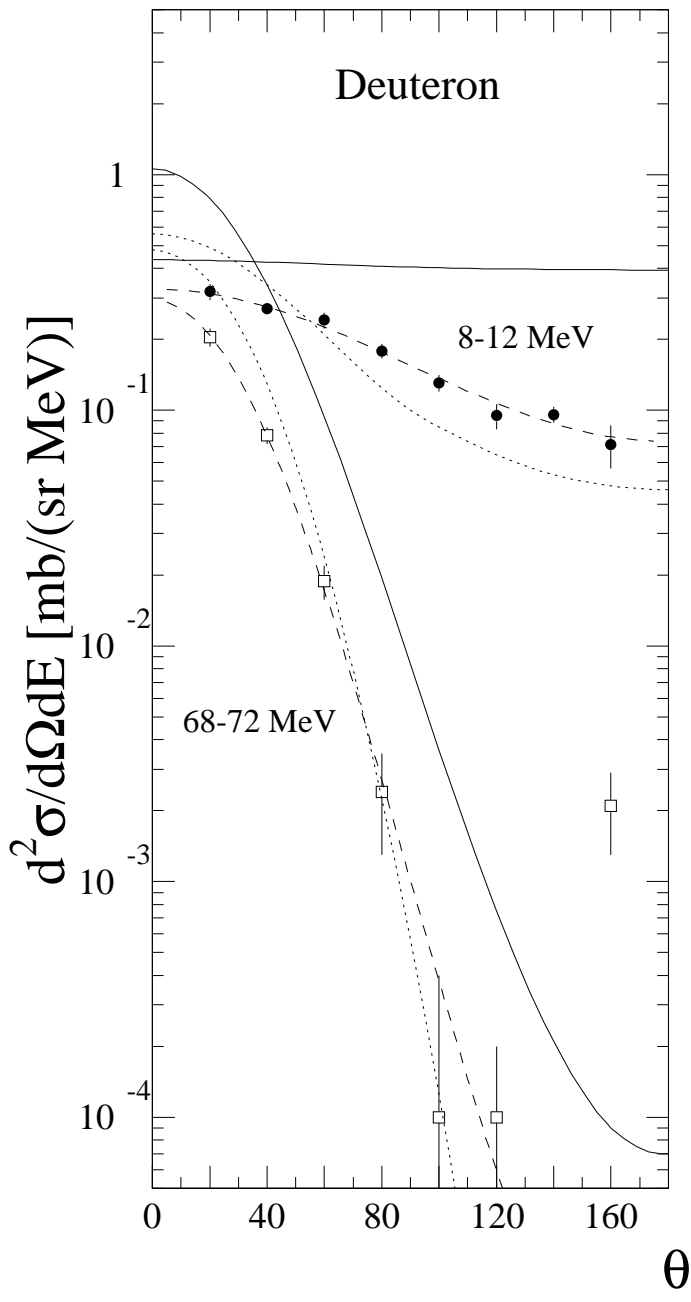
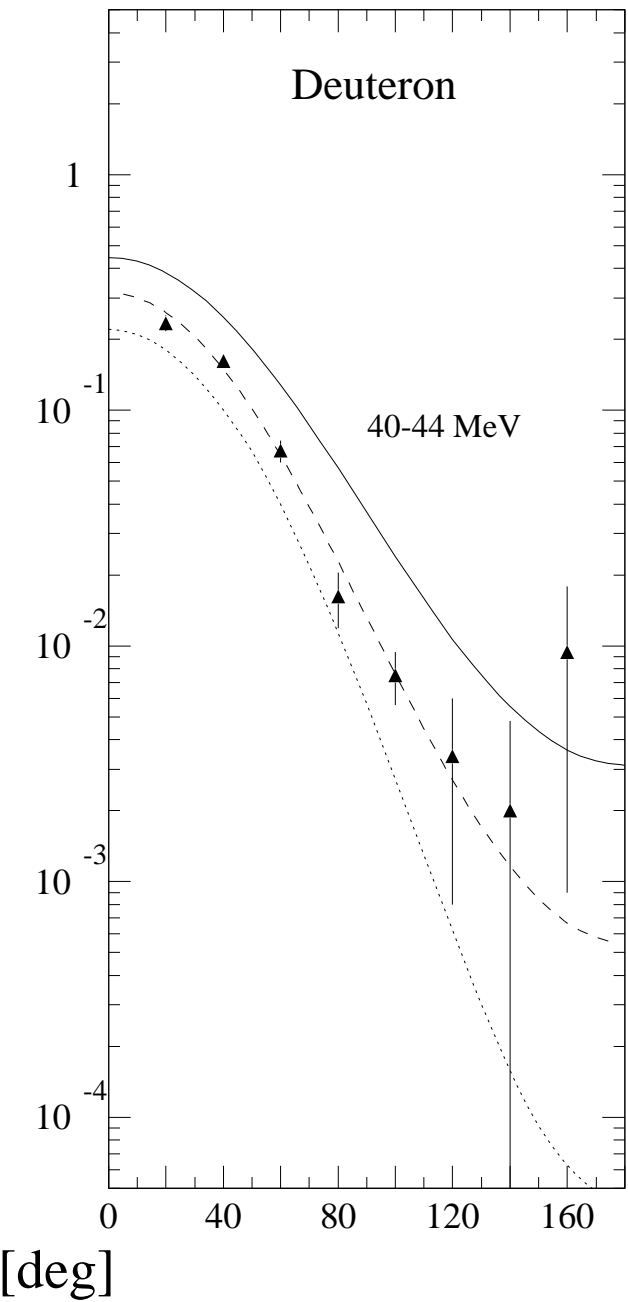
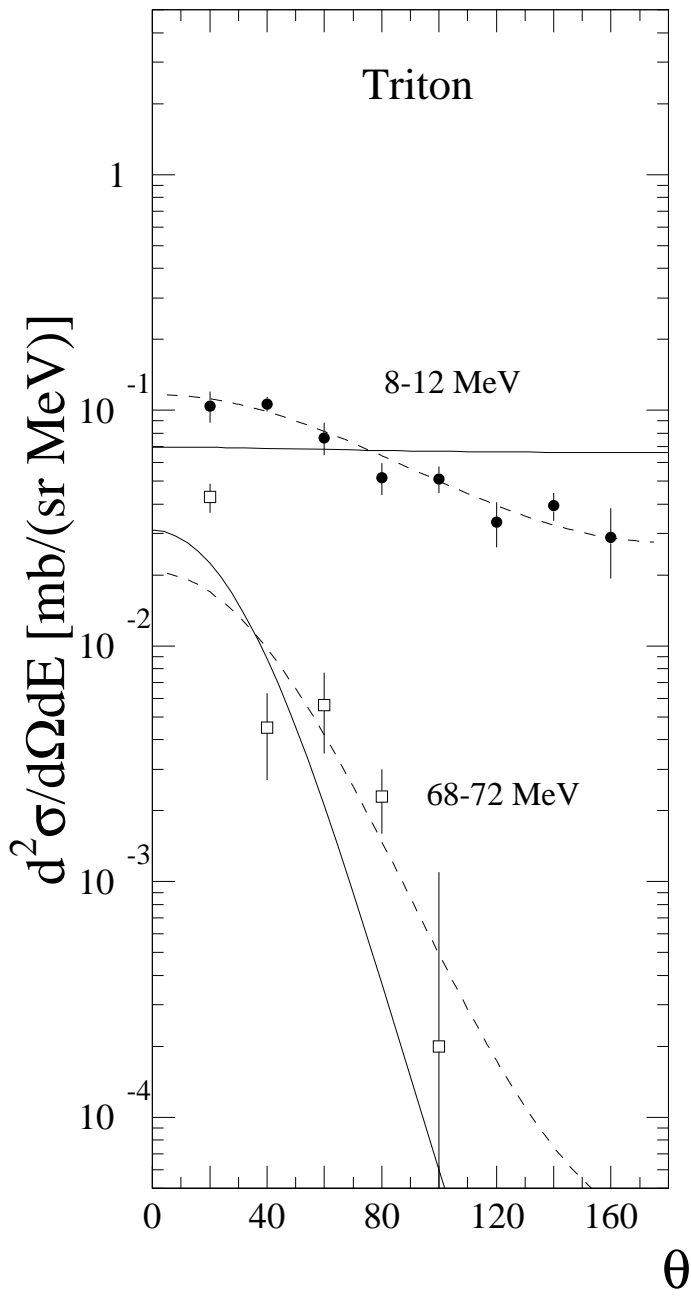
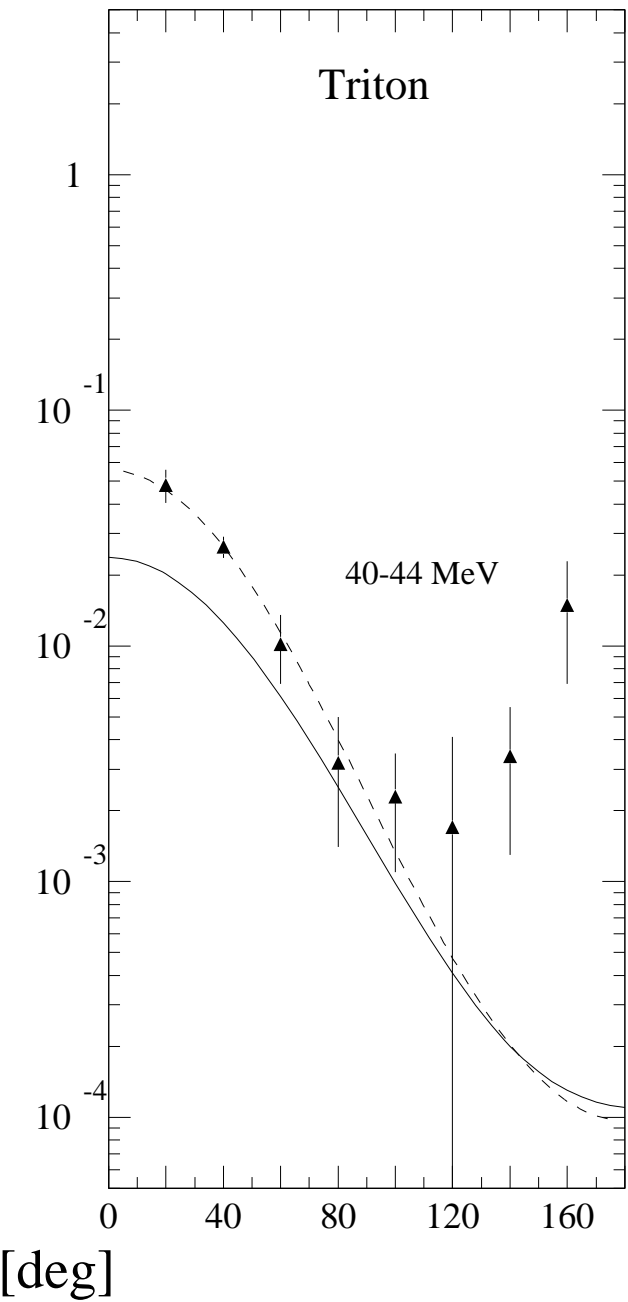
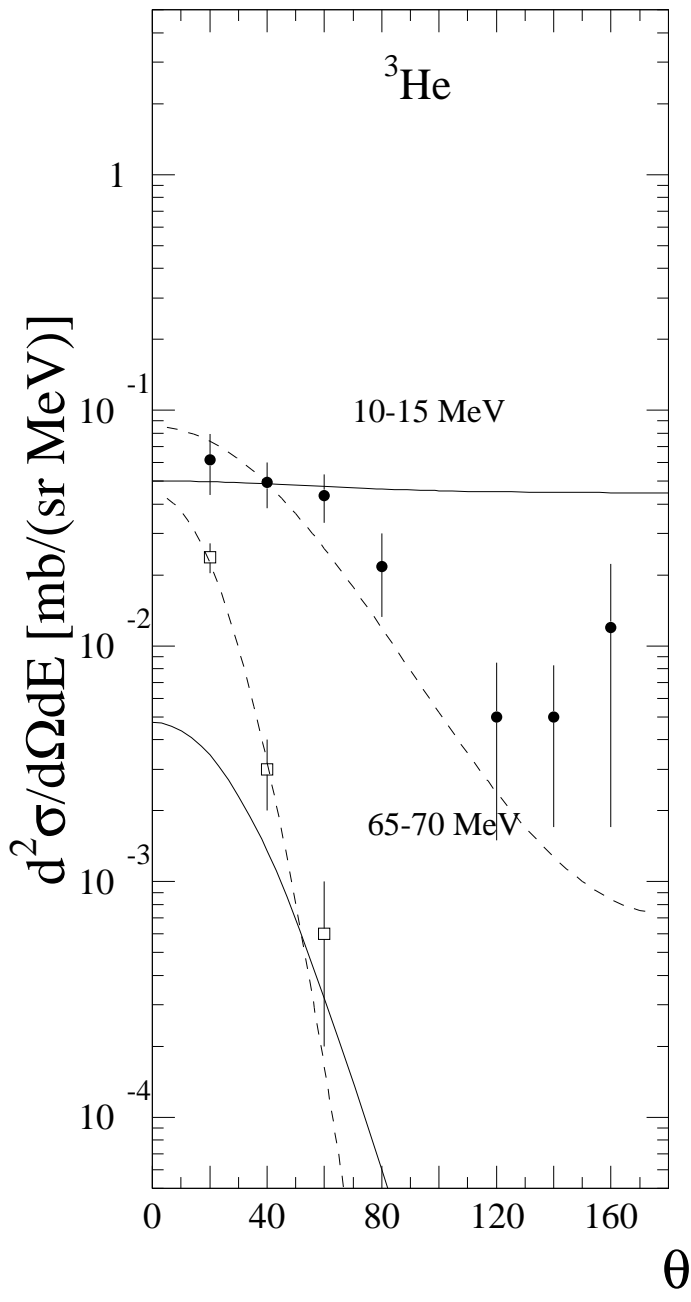
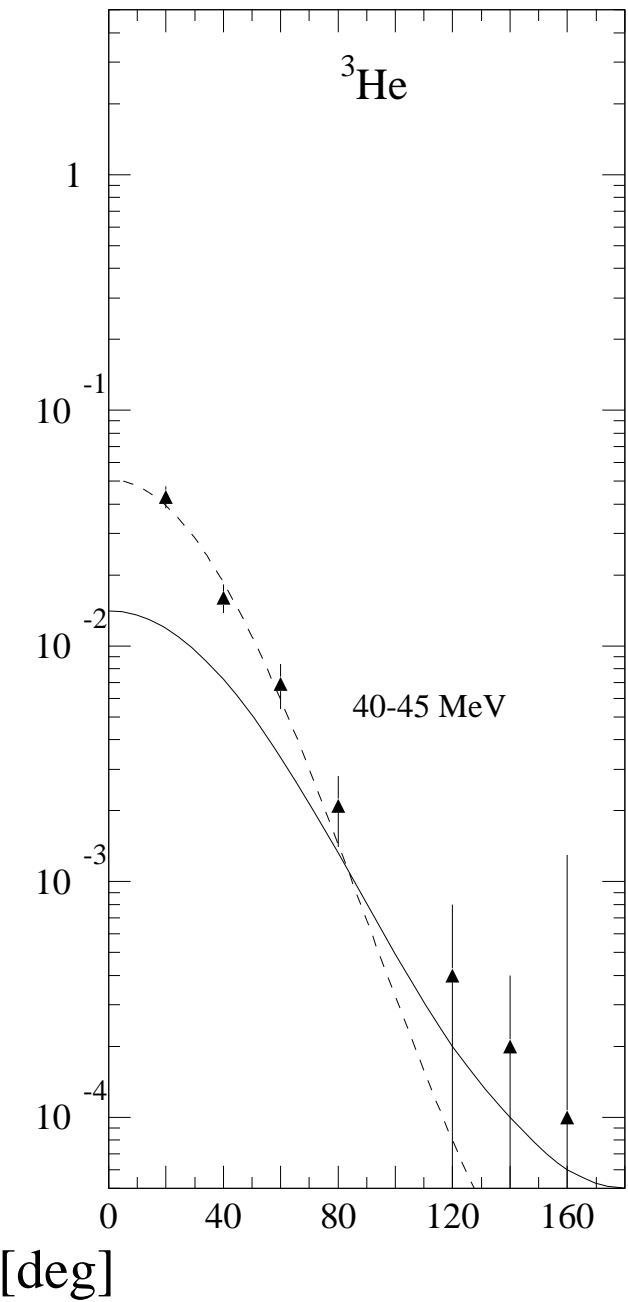
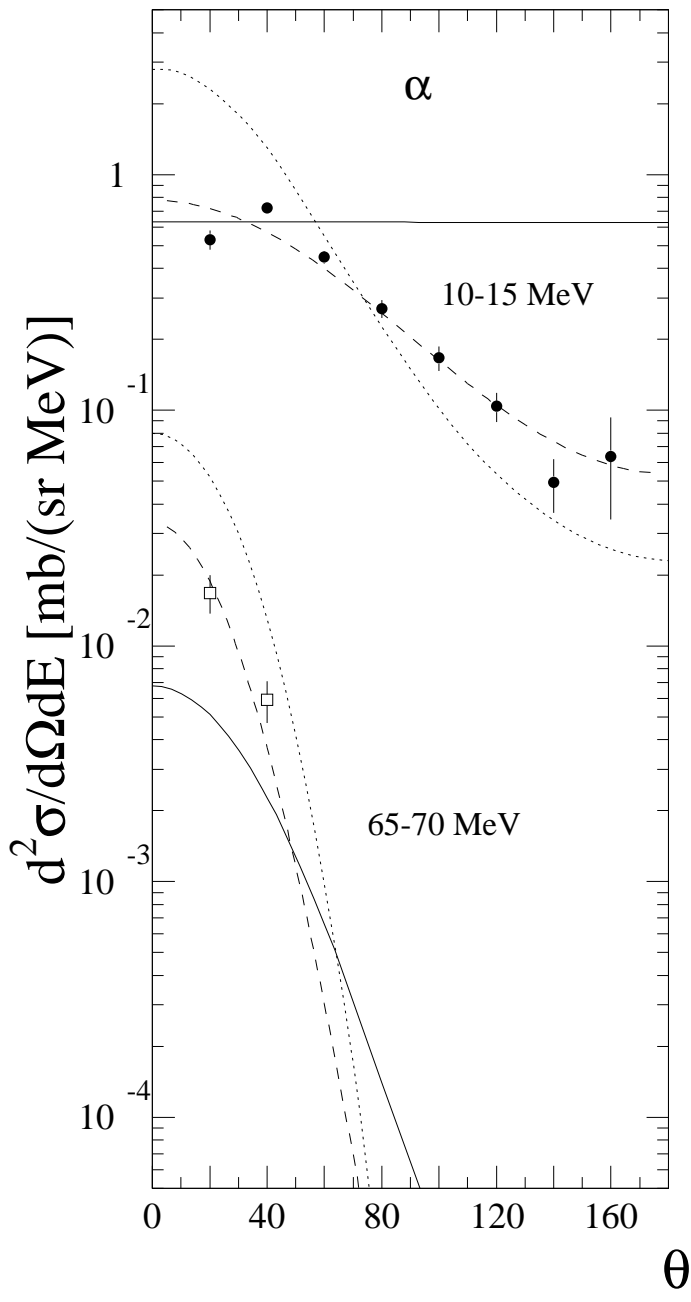
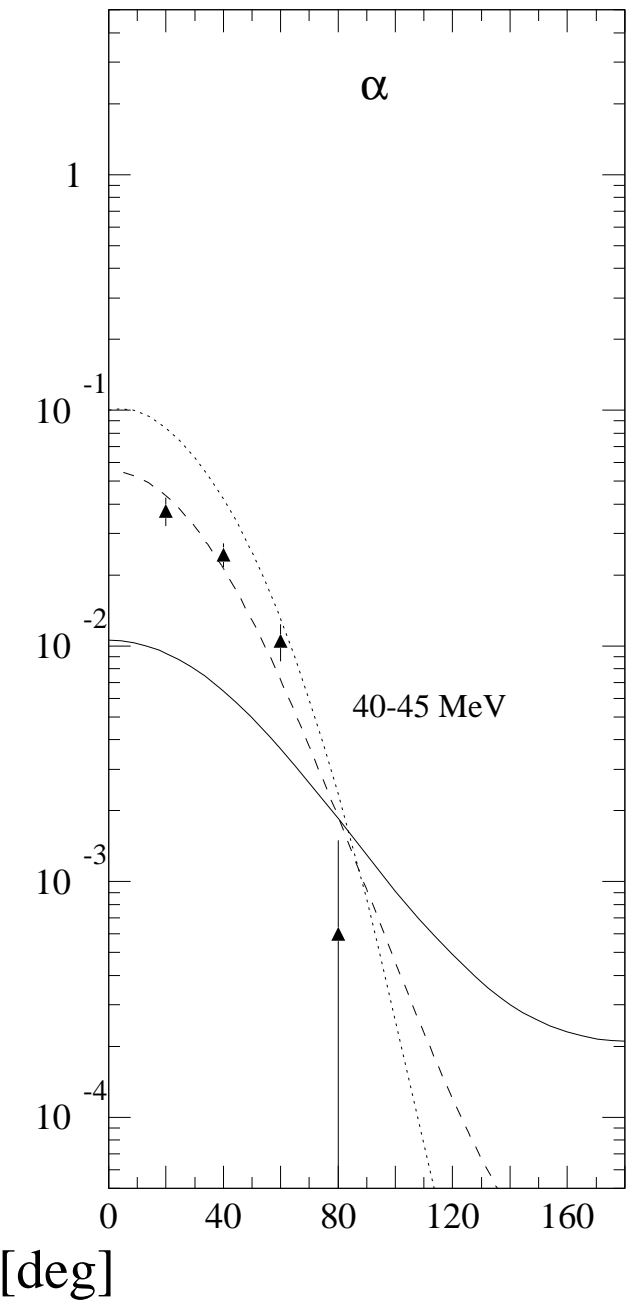
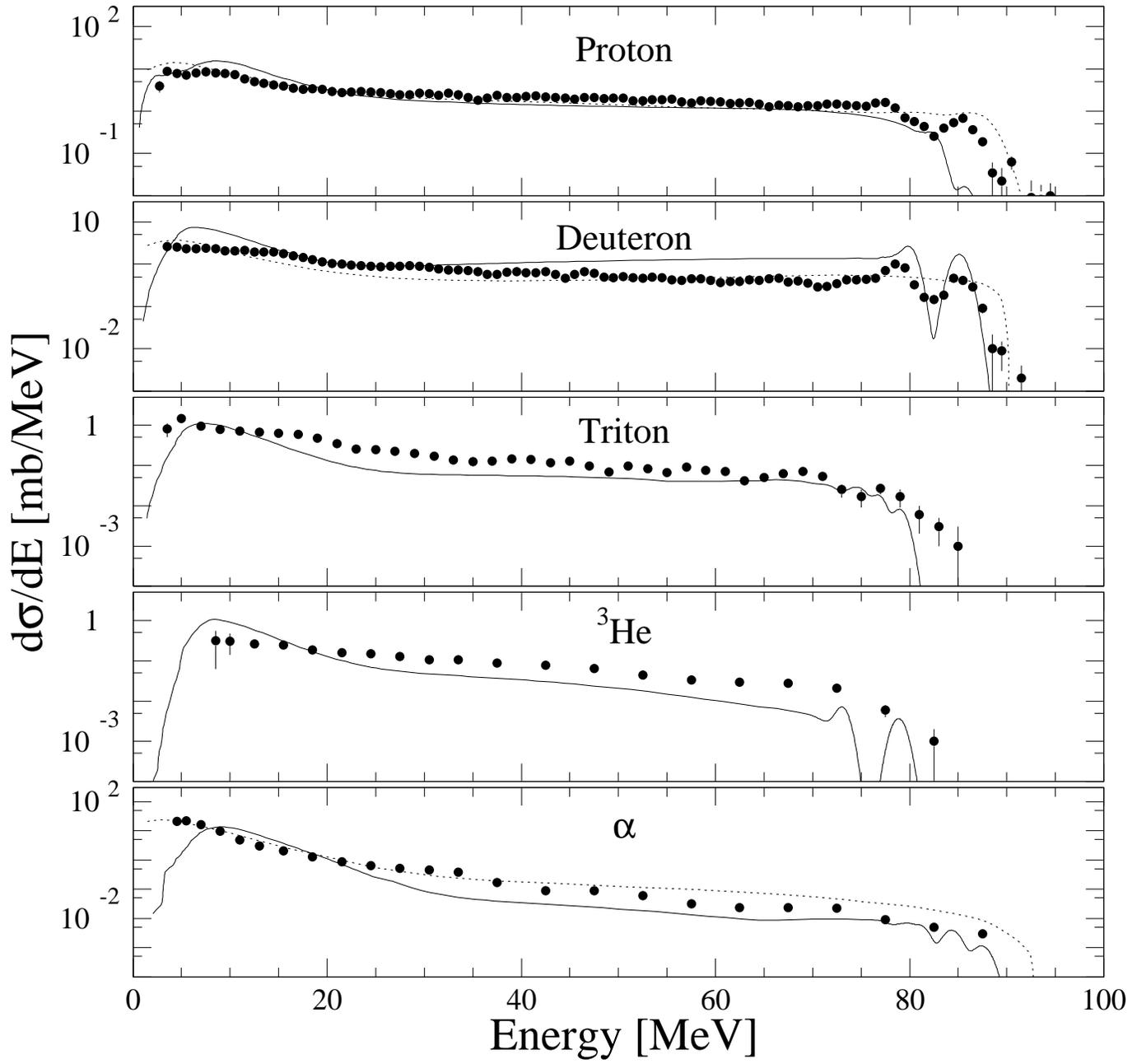
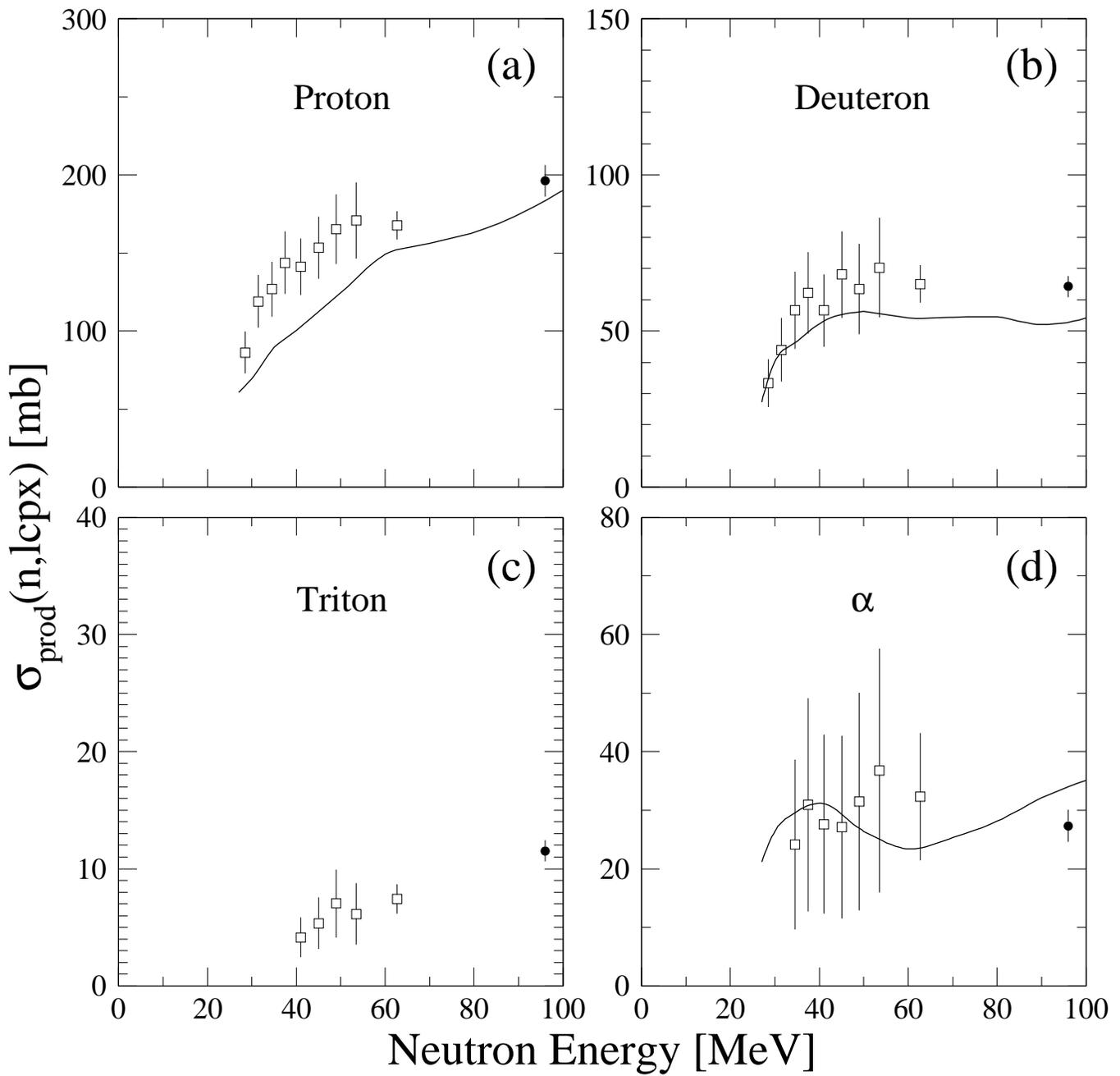
\end{document}